\documentclass{aa}
\usepackage{graphicx}
\begin{document}
\title{Warm gas in central regions of nearby galaxies}
\subtitle{Extended mapping of CO(3--2) emission}
\author{M. Dumke\inst{1,2,3}
\and    Ch.\ Nieten\inst{2}
\and    G. Thuma\inst{2}
\and    R. Wielebinski\inst{2}
\and    W. Walsh\inst{2}
}
\institute{
	SMTO, Steward Observatory, The University of Arizona,
	933 N.~Cherry Avenue, Tucson, Arizona 85721, USA
\and	Max-Planck-Institut f\"ur Radioastronomie,
	Auf dem H\"ugel 69, 53121 Bonn, Germany
\and    Institut de Radio Astronomie Millim\'etrique,
        300 Rue de la Piscine, 38406 St.~Martin d'H\`eres, France
	}
\offprints{M.D. (mdumke@as.arizona.edu)}
\date{Received 30 August 2000, Accepted 10 May 2001}
\titlerunning{CO(3--2) mapping of nearby galaxies}
\authorrunning{M. Dumke et al.}

\maketitle

\abstract{
We have mapped the CO(3--2) line emission from several nearby galaxies,
using the Heinrich-Hertz-Telescope on Mt.\ Graham, Arizona. Unlike earlier
observations, our investigation is not restricted to starburst galaxies,
but includes twelve galaxies of various types and in different stages of
star forming activity. Furthermore, we have not only observed the central
positions of these objects, but have obtained maps of the extended
CO(3--2) emission, with a typical map extent of 2 to 3 arcminutes in each
direction. Our
observations show that this extended mapping is necessary to reveal spatial
changes of the ISM properties within the galaxies.\\
In this paper we present the data sets and some data analysis. We compare
the galaxies in view of their morphology and excitation conditions, using
line ratios, luminosities and other properties, like the extent of the
CO(3--2) emission. The main results of this CO(3--2) survey are:\\
1.~In none of the observed objects the emission is confined to the nucleus,
as claimed in some earlier publications. CO(3--2) emission can be detected
for some objects to the same extent as the CO(2--1) and the CO(1--0) lines.\\
2.~The emission is more concentrated to the vicinity of star forming
structures (nuclear regions and spiral arms) than the lower CO transitions
for most of the observed objects. This is shown by decreasing
(3--2)/(1--0) line intensity ratios from the very centres towards larger
radii. The (deconvolved) sizes of the central emission peaks in the
CO(3--2) line vary from about 300\,pc up to 3\,kpc.\\
3.~The CO(3--2) luminosity is stronger in objects that contain
a nuclear starburst or morphological peculiarities. The total power emitted
in the CO(3--2) line from the central regions (i.e.\ excluding spiral
arms/outer disk) is highest in the starburst galaxies NGC\,2146, M\,82,
NGC\,3628, and in the spiral galaxy M\,51. When comparing the total power
normalized to the size of the emission region, the starburst galaxies
M\,82 and NGC\,253 show the highest values (about three times higher
than most other objects), while NGC\,278 and NGC\,4631 show the lowest.\\
4.~With the present spatial resolution, the line ratios $R_{3,1}$ seem to
be independent of Hubble type, color or luminosity. Most galaxies with
enhanced central star formation (``starbursts'') show line ratios
of the integrated intensities of
$R_{3,1} \sim 1.3$ in the very centre and $\sim 1.0$ at a radius
of about 1\,kpc. Objects with a ring-like (or double-peak if seen
edge-on) molecular gas distribution (NGC\,253, M\,82, and NGC\,4631)
show lower ratios.
The two galaxies that have CO(3--2) emission distributed over their spiral
arms (NGC\,891 and M\,51) show very low line ratios despite their high
infrared luminosities. This result suggests that CO emission in these
objects reflects a large amount of molecular gas, but not enhanced star
forming activity.\\
5.~Starburst galaxies show CO(3--2) emission also in their disks.
The line intensities are higher than that of normal galaxies. This suggests
that even if a starburst is a localized phenomenon, it is related to
different properties of the molecular gas over the whole galaxy.
}

\section{Introduction}

Lower lying CO line transitions which are observable in the
mm-wavelength range are a widely used tracer of molecular
hydrogen. Extensive work has been done for the CO(1--0) and the
CO(2--1) line, whose upper levels are lying 5.5\,K and 17\,K above the
ground level. Even if these observations are a reliable
tracer of the total column densities of H$_2$ in external galaxies,
where in general several molecular cloud complexes are covered by one
telescope beam, these lines cannot trace CO gas which is highly excited.
This higher excitation could be caused by collisional processes in a dense
medium or by a strong radiation field leading to higher temperatures.
Since these conditions often occur in the inner regions of molecular clouds
where star formation is taking place, other tracers are needed to determine
the physical state of the gas that actually is forming stars.

The $J=3$ level of the CO molecule lies 33\,K above the ground state,
and the critical density for the 3--2 transition is
$\sim 5\,10^4\,{\rm cm}^{-3}$.
It is therefore only excited if the gas is sufficiently warm and/or
dense. Whereas the lower transitions trace the large bulk of molecular
gas at low and moderate temperatures, the CO(3--2) line gives some
insight into the molecular gas in the immediate surroundings of star forming
regions, where gas properties may be different.

While numerous publications exist about observations of the
CO(1--0) and the CO(2--1) lines in nearby galaxies -- large surveys
in which only one or a few points in many galaxies have been observed
(e.g. Braine et al.\ \cite{braine+93}; Young et al.\ \cite{young+95})
as well as detailed mapping observations of selected objects
(e.g.\ Reuter et al.\ \cite{reuter+91}; Garc\'{\i}a-Burillo et
al.\ \cite{garcia+92}; Golla \& Wielebinski \cite{golla+rw94};
and many others) -- the number of published CO(3--2) observations is
still small. The reason for this was a lack of good sub-mm telescopes
on sites which allow observations at 345\,GHz, the rest frequency of
the CO(3--2) line. The first detection of extragalactic CO(3--2) was
reported by Ho et al.\ (\cite{ho+87}), in the nearby galaxy IC\,342.

In the following years, a few -- mainly starburst -- galaxies
have been investigated in the CO(3--2) line, but these observations
usually covered only the very centre. Recently Mauersberger
et al.\ (\cite{mauersberger+99}) observed a larger sample of about 30
nearby galaxies, but this was restricted to one position for
each of these objects. Extended maps of three nearby galaxies
were presented by Wielebinski et al.\ (\cite{rw+md+cn99}).

In this paper we report the extensive mapping of 12 large nearby
galaxies in the CO(3--2) line. The main
goal of these observations was to obtain a homogeneous
set of CO(3--2) data from a galaxy sample which covers various
Hubble types as well as different stages of activity.
While the spatial resolution of our observations is not sufficient to
investigate the physical conditions in individual
molecular clouds, the data provide information on the large-scale
distribution of highly excited CO gas in galaxies. Together with published
data from lower transitions it is possible to determine the global
excitation state of the molecular gas as a function of star forming
activity and the location of the emitting region in the
observed objects.

The main criterion used to select the target galaxies was the strength
of the CO(1--0) line. Further criteria were the expected global line-width
(which should be small enough to fit into the limited bandwidth) and the
availability of sufficiently large data sets of the lower CO transitions.
In addition the visibility at the observatory site was considered for the
selection.  In Sect.\ \ref{section:obs} we describe the performed
observations and the steps of reducing and calibrating the data,
while in Sect.\ \ref{section:results} we present the main results.
This section includes also a set of figures for each object. We compare
the observed galaxies in view of their morphology, kinematics, and
excitation conditions in Sect.\ \ref{section:discussion}, where also
some conclusions following from this work are presented.
The basic data obtained from this survey are collected in
Table \ref{tab:results}, which can also be found in
Sect.\ \ref{section:discussion}.

A more detailed analysis of the physical properties in the ISM of the
observed galaxies is beyond the scope of this paper. On some of the
observed objects more detailed investigations will be conducted in
forthcoming publications.

\section{Observations and data reduction}
\label{section:obs}

\begin{table}
\caption[]{Galaxies mapped in the CO(3--2) emission, with their adopted
positions and LSR velocities.
The coordinates given here represent
the (0,0) position in the maps shown in Sect.\ \ref{section:results}.
References are:
NGC\,253, NGC\,278, NGC\,2146, NGC\,4631, and NGC\,6946
(Young et al.\ \cite{young+95}),
NGC\,891 (Garc\'{\i}a-Burillo et al.\ \cite{garcia+92}),
Maffei\,2 (Weliachew et al.\ \cite{weliachew+88}),
IC\,342 (Ishizuki et al.\ \cite{ishizuki+90}),
M\,82 (Rieke et al.\ \cite{rieke+80}),
NGC\,3628 (Irwin \& Sofue \cite{irwin+sofue96}),
M\,51 (Garc\'{\i}a-Burillo et al.\ \cite{garcia+93}),
and M\,83 (Petitpas \& Wilson \cite{petitpas+wilson98}).
}
\label{tab:obs}
\begin{tabular}{lrrc}
\hline
Source & R.A.(2000) & Dec.(2000) & $v_{\rm lsr}$ [km\,s$^{-1}$]
	\rule{0pt}{9pt}\\
\hline
NGC\,253 & $0^{\rm h}47^{\rm m}33^{\rm s}\!\!.0$ & $-25^{\circ}17'20''$ &
	251 \\
NGC\,278 & $0^{\rm h}52^{\rm m}04^{\rm s}\!\!.5$ & $47^{\circ}33'01''$ &
	640 \\
NGC\,891 & $2^{\rm h}22^{\rm m}33^{\rm s}\!\!.1$ & $42^{\circ}20'55''$ &
	527 \\
Maffei 2 & $2^{\rm h}41^{\rm m}55^{\rm s}\!\!.2$ & $59^{\circ}36'11''$ &
	-36 \\
IC\,342 & $3^{\rm h}46^{\rm m}48^{\rm s}\!\!.3$ & $68^{\circ}05'46''$ &
	32 \\
NGC\,2146 & $6^{\rm h}18^{\rm m}37^{\rm s}\!\!.6$ & $78^{\circ}21'19''$ &
	903 \\
M\,82 & $9^{\rm h}55^{\rm m}52^{\rm s}\!\!.6$ & $69^{\circ}40'47''$ &
	225 \\
NGC\,3628 & $11^{\rm h}20^{\rm m}17^{\rm s}\!\!.0$ & $13^{\circ}35'20''$ &
	844 \\
NGC\,4631 & $12^{\rm h}42^{\rm m}07^{\rm s}\!\!.7$ & $32^{\circ}32'28''$ &
	608 \\
M\,51 & $13^{\rm h}29^{\rm m}52^{\rm s}\!\!.5$ & $47^{\circ}11'46''$ &
	470 \\
M\,83 & $13^{\rm h}37^{\rm m}00^{\rm s}\!\!.7$ & $-29^{\circ}51'58''$ &
	510 \\
NGC\,6946 & $20^{\rm h}34^{\rm m}51^{\rm s}\!\!.9$ & $60^{\circ}09'15''$ &
	46 \\
\hline
\end{tabular}
\end{table}

We used the Heinrich-Hertz-Telescope\footnote[1]{The HHT is operated by the
Submillimeter Telescope Observatory on behalf of Steward Observatory
and the MPI f\"ur Radioastronomie.}
(Baars \& Martin \cite{baars+martin96}; Baars et al.\ \cite{baars+99}),
located on Mt.\ Graham, Arizona, during four periods of variable
length in April 1998, December 1998, April 1999, and January 2000,
to obtain the data presented in this paper.

We usually had several hours of good atmospheric conditions with a zenith
opacity between 0.2 and 0.8 at 345\,GHz. For the observations we used a
2-channel SIS receiver (with the channels sensitive to orthogonal
polarizations), provided by the MPIfR Bonn. Each of the two receiver
channels was connected to a 2048-channel AOS, providing a total bandwidth
of 1\,GHz and a channel separation of 0.48\,MHz
(velocity resolution $\sim 0.9\,{\rm km\,s}^{-1}$).
System temperatures for both receiver channels were typically between
600 and 1200\,K.

All results in this paper are given in main beam brightness temperatures
($T_{\rm mb}$), which are related to the antenna temperature, corrected for
atmospheric absorption and rear spillover via
$T_{\rm mb} = (F_{\rm eff}/B_{\rm eff}) T_{\rm A}^*$ (the IRAM calibration
scheme). The main beam efficiency of the HHT at 345\,GHz is about 0.46 and
the forward efficiency about 0.92 (Wilson et al.\ \cite{wilson+00}).

We used the wobbler-switch mode for our observations, with the subreflector
wobbling with a frequency of 2\,Hz and the reference positions located at
$\pm 4'$. All target observations were made on a rectangular grid in
RA-Dec-orientation with a pixel separation of $10''$ in each direction,
in order to achieve a full sampling for the given beamwidth of about
$22''$. We regularly observed strong continuum sources (planets, W3OH)
and also point-like spectral line sources to check the pointing accuracy.
We also checked the focussing
of the telescope two times a day on a planet. While the focus parameters
were stable within $\pm 0.03\,{\rm mm}$ and the pointing accuracy
was usually about $3''$, the observations of some days in April 1998
suffered from larger pointing errors (up to $15''$) due to an encoder problem.
We could, however, overcome this problem by regularly mapping the inner
parts of the target galaxies, where line shapes change rapidly. This allowed
an accurate relative positioning of the individual observations of each galaxy.
An absolute pointing accuracy of $5''$ (maximum errors) was finally obtained.
This was also confirmed by re-observing every source after the encoder
problem was fixed. This pointing problem has only a negligible effect on
our absolute calibration (because our sources are not point-like).
But together with a small misalignment of both receiver channels due to
the complicated optics of the receiver it may degrade the beam width of
the reduced and co-added spectra from nominal $22''$ (HPBW) to about $24''$.

The temperature scale calibration procedure is described by
Mauersberger et al.\ (\cite{mauersberger+99}); the absolute calibration
was checked by regular observations of galactic spectral line sources.
We did not find any systematic deviations of our results from the
brightness temperatures listed in Stanek et al.\ (\cite{stanek+95}).
However, some additional calibration uncertainty can arise from a
sideband ratio $\neq 1$ with the used DSB receiver, and from the fact
that the sideband ratio may have varied for individual tunings. Hence
the {\it absolute} calibration scale may be accurate to 15\,\%, what is
not unusual in the sub-mm range. The {\it relative} calibration scale
could be checked by re-observing the central region of each object for
every individual tuning. After adjusting the temperature scale of some
data sets, the remaining relative calibration uncertainty is
$< 3\,\%$ for objects with narrow lines (like NGC\,278 and IC\,342)
and may be up to 8\,\% over the whole mapped region for objects which
cover a large velocity range, or between various objects at different
velocities. The error estimates given in Sect.\ \ref{section:results}
take into account only the relative uncertainties (besides the spectral
noise), since only these are important for a comparison between the
objects and various regions within one object.

The data reduction was performed using the GILDAS-package including the
CLASS and GRAPHIC programs in a standard manner. Baselines of first order
were subtracted from each spectrum. In many cases the baselines
of receiver channel 2 were curved; from these spectra we subracted also
baselines of second or third order and checked the results by comparison
with the corresponding spectra of channel 1. Finally several individual
spectra on each position of the observed region were coadded with an
appropriate weighting.

\section{Results}
\label{section:results}

The target galaxies selected for this study and for which we
have obtained extended maps are listed
in Table \ref{tab:obs}. Some important data from the
literature are compiled in Table \ref{tab:basic}, while our results are
collected in Tab.\ \ref{tab:results}.
These include the measured line flux (i.e.\ the velocity-integrated line
intensity, spatially integrated over the central emission peak), the power
radiated in the CO(3--2) line (also spatially integrated),
the spatial extent of the central emission peak, the emitted power normalized
to the size of the emitting region, line ratios in the centre and the disk,
and the line width in the centre, which is {\em not} corrected for different
inclinations.
Both Tables \ref{tab:basic} and
\ref{tab:results} can be found in Sect.\ \ref{section:discussion}.
In this section we will refer to line ratios $R$\/, total fluxes $F$\/,
and other terms; their definitions can also be found in
Sect.\ \ref{section:discussion}.

For each galaxy we present a set of figures illustrating the basic results.
These are (a) a raster map of the spectra we obtained within a galaxy,
(b) a contour map of the velocity-integrated line intensity,
overlaid on an optical image extracted from the
Digitized Sky Survey\footnote[2]{The DSS images are based on photographic
data obtained using Oschin Schmidt Telescope on Palomar Mountain.
The Digitized Sky Survey was produced at the Space Telescope Science
Institute under U.S. Government grant NAG W-2166.},
(c) a position-velocity diagram along the major axis which is defined by
the position angle (ccw from north) in Table \ref{tab:basic}, and
(d) a small graph\footnote[3]{We acknowledge the use of the GRACE program
for plotting purposes (http://plasma-gate.weizmann.ac.il/Grace/).}
showing the measured variation with galactocentric distance of
the velocity-integrated intensity $I_{\rm CO(3-2)}$,
both along the major and minor axis.
While the raster maps (a) are shown with velocity re\-so\-lu\-tions of
$\Delta v = 10\,{\rm km\,s}^{-1}$ in the central regions and
20\,km\,s$^{-1}$ in the outer parts for some galaxies, the other figures
and the data analysis are done using data
cubes with $\Delta v = 10\,{\rm km\,s}^{-1}$ for all positions.
For the position-velocity diagrams (c)
and the intensity distribution (d) the offsets along the axes are always
chosen in a way that increasing offsets correspond to increasing declinations.
This means that the strips go either from southeast to northwest or from
southwest to northeast, even if this seems to be in contradiction with the
usual definition of the position angle, which is counted up to $180^{\circ}$
counterclockwise from north. Further we note that for some of the galaxies
(NGC\,253, Maffei\,2, IC\,342, NGC\,2146, M\,82, and NGC\,4631)
the zero positions of figures (c) and (d) do not coincide with the
central positions given in Table \ref{tab:obs}, but are shifted to the
CO(3--2) maxima; see the corresponding subsections for details.
This was done in cases where a wrong reference position for the cuts
along the major and minor axis could have lead to incorrect results.
In figure parts (a) and (b), however, the (0,0) position always
corresponds to that given in Table \ref{tab:obs}.

The spectra of three of the galaxies (NGC\,278, NGC\,4631, and M\,51) have
already been published by Wielebinski et al.\ (\cite{rw+md+cn99});
nevertheless we have included these in order to have a complete
sample of galaxies which are analysed in the same manner. This provides
better statistics for our conclusions.

The rms noise in our spectra depends on the observing time spent on each
point. The latter was chosen according to the strength of the signal and
the location of the grid point within the galaxy. For the final velocity
resolution of 10\,km\,s$^{-1}$, near the centre of the mapped objects the
channel-to-channel rms is ranging from 20\,mK to about 60\,mK. Near map
edges the rms is generally increasing, reaching about 160\,mK at worst.
In general, the uncertainties due to spectral noise are smaller than the
overall calibration uncertainty as discussed in Sect.\ \ref{section:obs}.
We further note that any {\it absolute} calibration error is of a systematic
nature and has a similar effect for all objects in our sample, minimizing
the effect on the conclusions one can draw from a comparison of the galaxies.

\subsection{NGC\,253}
\label{section:results_n253}

\begin{figure*}
\resizebox{\hsize}{!}{\includegraphics*{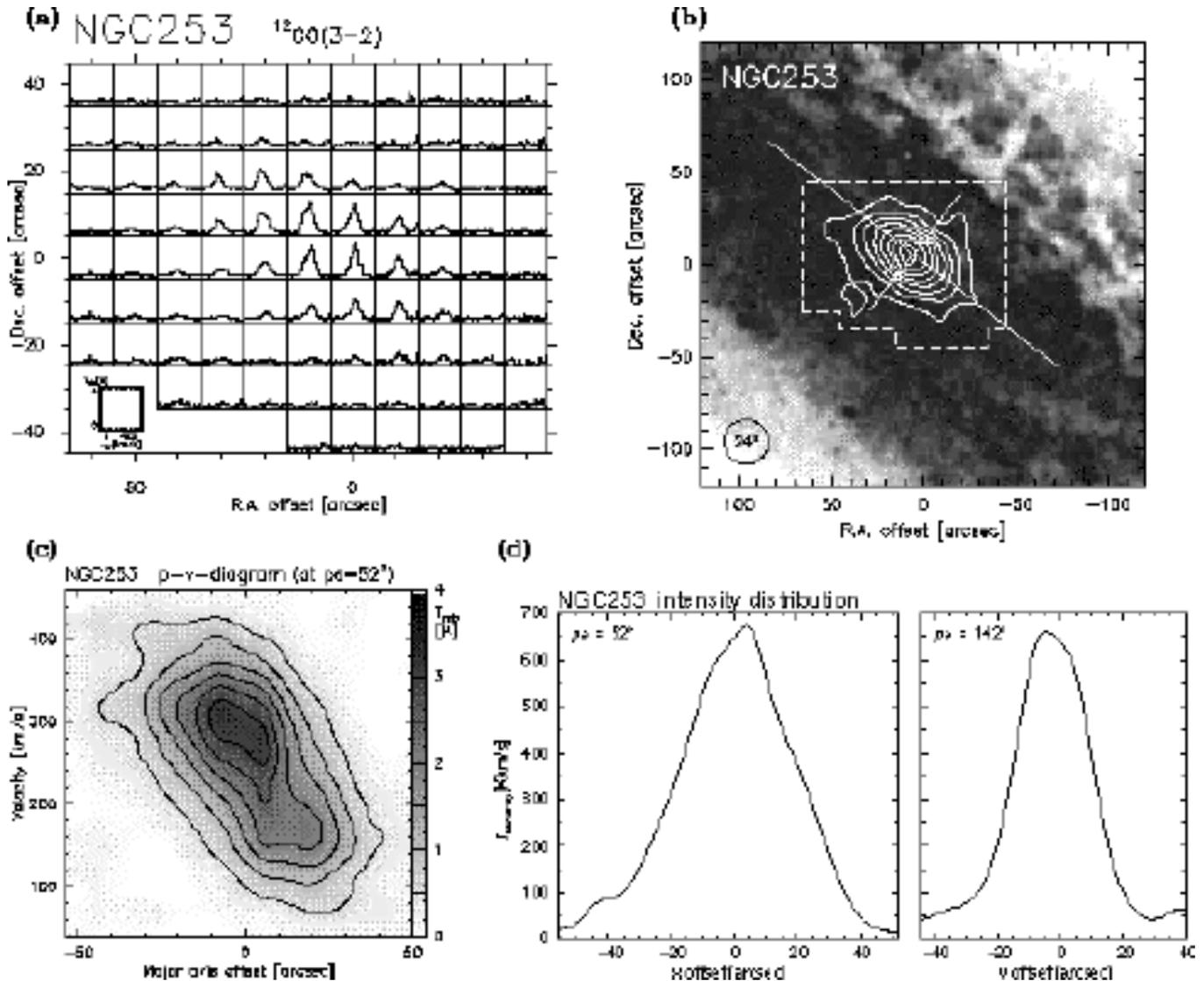}}
\caption{CO(3--2) data of NGC\,253:
{\bf a} (upper left) raster map of the individual spectra, with (0,0)
corresponding to the position given in Table \ref{tab:obs}.
The scale of the spectra is indicated by the small box inserted
in the lower left corner of the image.
{\bf b} (upper right) contour map of the integrated intensity, overlaid on an
optical image extracted from the Digitized Sky Survey. Contour levels are
80, 160, \ldots, 640\,K\,km\,s$^{-1}$.
The region covered by the spectra shown in (a) is indicated by the dashed
polygon, and the cross shows the reference position for data analysis
(see text) and the direction of the major and minor axis.
{\bf c} (lower left) position-velocity diagram along the major axis. Contours
are 0.5, 1.0, \ldots 3.0\,K.
{\bf d} (lower right) intensity distribution along the major and minor
axis, respectively. Note that for figure parts {\bf c} and {\bf d} the zero
position is taken from Mauersberger et al.\ (\cite{mauersberger+96}) and
differs from the position (0,0) in figure parts {\bf a} and {\bf b} by
$(\Delta {\rm R.A.}, \Delta{\rm Dec.}) = (5''\!\!.4, 6''\!\!.0)$ 
}
\label{fig:n253}
\end{figure*}

The Sculptor galaxy NGC\,253 is one of the nearest ($D = 2.5\,{\rm Mpc}$;
Mauersberger et al.\ (\cite{mauersberger+96}) and references therein)
spiral galaxies which exhibit strongly enhanced star formation
(a ``starburst'') in its central region (Rieke et al.\ \cite{rieke+88}).
It is nearly edge-on ($i = 78^{\circ}$, Pence \cite{pence81})
and is classified as SAB(s)c. The starburst nucleus of NGC\,253 is one
of the brightest extragalactic sources from the infrared to
cm-wavelengths. It is presumably triggered by the potential of a bar that
is visible in many wavelength ranges.

Due to its small distance and its luminosity this galaxy has already
been studied by several authors in detail. There are also numerous
investigations of the CO gas. Wall et al.\ (\cite{wall+91}) observed
NGC\,253 in the (2--1) and (3--2) lines of the isotopomers
\element[][12]{CO} and \element[][13]{CO}. They analysed the data to
determine excitation conditions for the major and minor axis.
Large maps of the CO(1--0) and the CO(2--1) emission of this galaxy were
obtained by Houghton et al.\ (\cite{houghton+97}) and
Mauersberger et al.\ (\cite{mauersberger+96}). The higher
transitions (3--2) and (4--3) of the CO molecule were also
observed, and maps are presented by Israel et al.\ (\cite{israel+95}).
The most recent CO study, including the three lowest transitions of the
most abundant isotopomers, was published by
Harrison et al.\ (\cite{harrison+99}).

{\it Morphology.}
The extent of this galaxy is several arc\-minutes, but
our observations are restricted to the inner part of the optical disk,
out to a radius of about $65''$ (800\,pc).
The data we obtained are shown in Fig.\ \ref{fig:n253}.
The highest brightness temperatures in the central region reach values
of 3 to 4\,K. Israel et al.\ (\cite{israel+95}) measured
somewhat higher temperatures ($5 - 6\,{\rm K}$). These are not in
contradiction with our values due to the smaller beam of the JCMT.
However, Wall et al.\ (\cite{wall+91}) also observed peak temperatures
of about 6\,K in the CO(3--2) line with the CSO, thus with a beam size
similar to our observations.

The central position of our observations was originally chosen
at the coordinates given by Young et al.\ (\cite{young+95}). It was shifted
afterwards to the value given in Table \ref{tab:obs} to better fit
to the central position often given by other authors. The intensity
peak of the CO(3--2) data, moreover, is located at the position given
by Mauersberger et al.\ (\cite{mauersberger+96}), which is at
$(\Delta {\rm R.A.}, \Delta{\rm Dec.}) = (5''\!\!.4, 6''\!\!.0)$
relative to the (0,0) position. This was taken as reference point
for the data analysis (Fig.\ \ref{fig:n253}c and d).

The morphology of the observed CO(3--2) emission is -- considering the
different beams -- similar to the maps given by Israel et
al.\ (\cite{israel+95}). We cannot resolve the two intensity peaks
(which are separated by $\sim 10''$) in the total intensity map, but
a twin-peak structure is suggested by the position-velocity diagram along
the major axis of NGC\,253, with the stronger peak in the southwest.
The source size can be estimated by Gaussian fits along the major and
minor axis. When deconvolved from the telescope beam, assumed to
be $24''$ (see Sect.\ \ref{section:obs}), this is
$35'' \times 15''$, which corresponds to a linear size of
$430\,{\rm pc} \times 180\,{\rm pc}$ at the distance of NGC\,253. These
values suggest that the molecular gas temperature and density fall off
relatively quickly with increasing distance from the centre, even along
the bar.

{\it Intensities.}
The integrated CO(3--2) flux of the region mapped in NGC 253 is
$F_{\rm CO(3-2)} = 8.74 \pm 0.30\,10^4\,{\rm Jy\,km\,s}^{-1}$. The total
power radiated in the CO(3--2) line is
$6.0 \pm 0.2\,10^{30}\,{\rm W}$. Since we have not mapped the whole galaxy,
but only the inner $\sim 2\,{\rm kpc}$ of the optical disk,
the total values (for the whole galaxy) are
probably somewhat larger. Houghton et al.\ (\cite{houghton+97}) have
detected CO(1--0) emission out to a radius of $11'$ (8\,kpc),
hence there may be also
an extended CO(3--2) component far away from the centre.

The maximum integrated line intensity in our map (Fig.\ \ref{fig:n253}b)
is $I_{\rm CO(3-2)} = 680 \pm 60\,{\rm K\,km\,s}^{-1}$. This value is
somewhat lower than data given by other authors, even when observed with (or
convolved to) similar telescope beams. Harrison et al.\ (\cite{harrison+99})
give a value of $998\,{\rm K\,km\,s}^{-1}$, which is consistent with
$990\,{\rm K\,km\,s}^{-1}$ given by Israel et al.\ (\cite{israel+95}).
Wall et al.\ (\cite{wall+91}) measured even $1200\,\rm{K\,km\,s}^{-1}$.
Hence it seems that our value for $I_{\rm CO(3-2)}$ is about 30\,\% below
the other values, and this difference is larger than our overall calibration
uncertainty. It is very unlikely that we have lost some emission due to
incorrect baseline subtraction. We should also note that data for this
galaxy were taken during different observing periods, and the individual
data sets were consistent with each other.
Hence there is no obvious explanation for the intensity differences
compared to earlier publications.
However, atmospheric calibration errors more severe than expected cannot
be completely ruled out, considering the low declination of NGC\,253 at
the HHT.

\begin{figure*}
\resizebox{\hsize}{!}{\includegraphics*{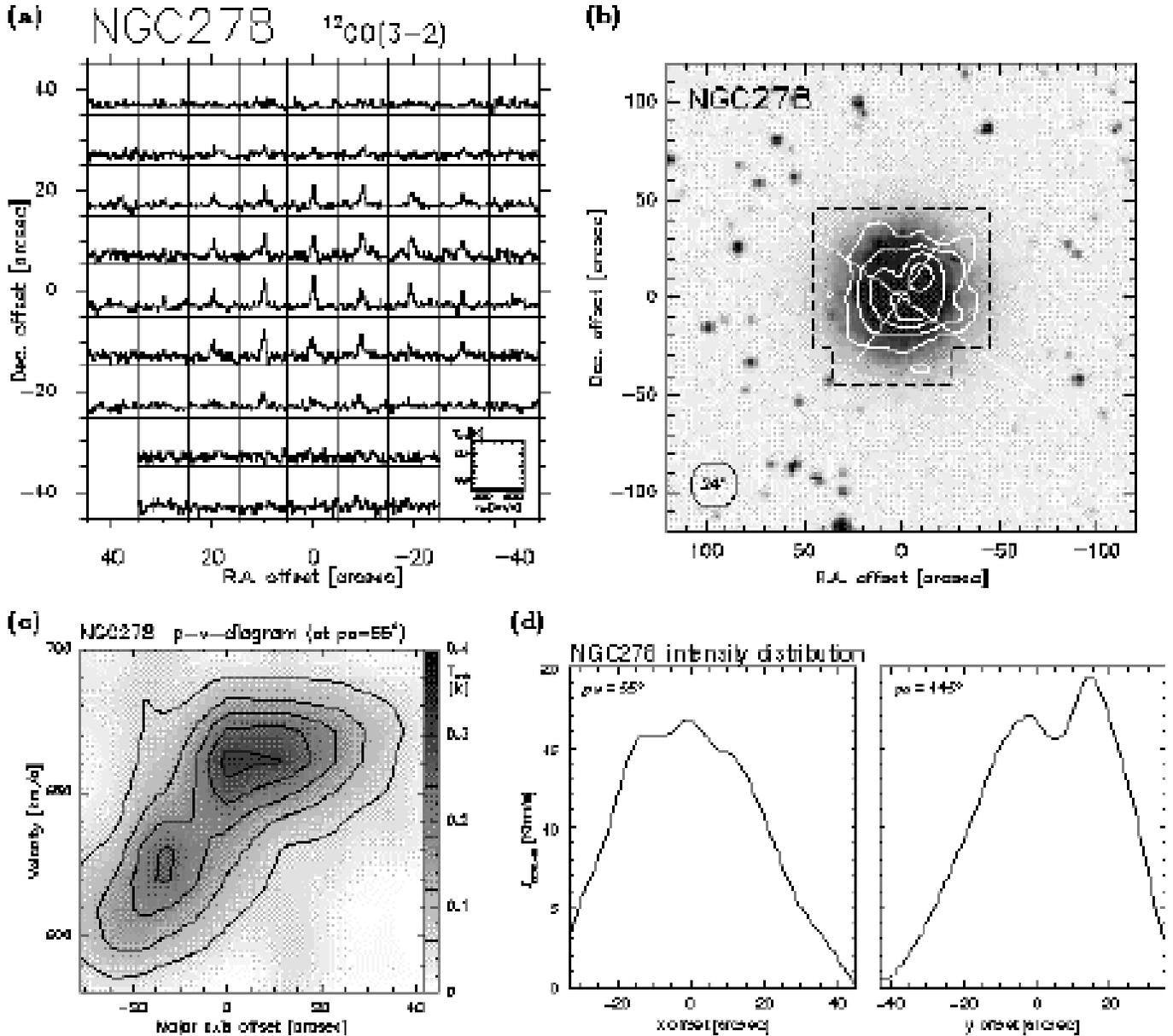}}
\caption{CO(3--2) data of NGC\,278:
{\bf a} (upper left) raster map of the individual spectra, with (0,0)
corresponding to the position given in Table \ref{tab:obs}.
The scale of the spectra is indicated by the small box inserted
in the lower right corner of the image.
{\bf b} (upper right) contour map of the integrated intensity, overlaid on an
optical image extracted from the Digitized Sky Survey. Contour levels are
4, 8, 12, 16\,K\,km\,s$^{-1}$.
The region covered by the spectra shown in (a) is indicated by the dashed
polygon, and the cross shows the adopted centre and the direction
of the major and minor axis.
{\bf c} (lower left) position-velocity diagram along the major axis. Contours
are 0.06, 0.12, \ldots 0.30\,K.
{\bf d} (lower right) intensity distribution along the major and minor
axis, respectively
}
\label{fig:n278}
\end{figure*}

{\it Line Ratios.}
Because of this calibration difference,
our estimated line ratios, using the CO(1--0) and the CO(2--1) data
from Wall et al.\ (\cite{wall+91}) and Harrison et al.\ (\cite{harrison+99}),
are systematically below those estimated by these authors.
We find $R_{3,1} = 0.8 \pm 0.1$ and $R_{3,2} = 0.7 \pm 0.1$ in the
central region. For the disk, at a distance at about $30''$ from the
centre, we get a slightly lower value, $R_{3,2} = 0.5 \pm 0.1$. This
decrease in the line ratio can also be seen in the data of 
Wall et al.\ (\cite{wall+91}). However, the value for the central region
is surprisingly low when compared with those for other starburst galaxies
in our sample (see Table \ref{tab:results}). When we assume a CO(3--2)
line intensity 30\,\% higher, similar to the value of Harrison et
al.\ (\cite{harrison+99}), we would estimate $R_{3,1} \sim 1.1$ for
the central region. This is still below the average value for the other
starburst galaxies. A low line ratio is even more surprising when we consider
that, due to the small distance to NGC\,253, we sample only the very inner
star-forming centre, whereas for other galaxies a larger part of the
disk is also observed on the central points due to the lower linear
resolution. These line ratios indicate densities of
$10^3 - 10^{3.5}\,{\rm cm}^{-3}$ at temperatures of $20 - 40\,{\rm K}$
in the centre, but $n({\rm H}_2) < 10^3\,{\rm cm}^{-3}$ in the bar.
As for all of the observed galaxies, the values given for kinetic
temperature and density are only approximate, and a more detailed
discussion is only possible with the use of $^{13}{\rm CO}$ data.
For NGC\,253, we also refer to the publications cited above.

{\it Kinematics.}
We see only the rigidly rotating part of this galaxy in the CO(3--2) line. 
The total linewidth of the CO(3--2) emission at the position (10,10) was
determined by a Gaussian fit to be $\Delta\,V \sim 200\,{\rm km\,s}^{-1}$.
However, in the central region the velocity distribution of the emission
can be described by two (or even three) components, corresponding to the
double peak structure mentioned above, with a line width of about
120\,km\,s$^{-1}$ for each component.

\subsection{NGC\,278}

NGC\,278 is a small ($D_{25} = 2'\!\!.1$), barred Sb galaxy.
Its distance is 12.4\,Mpc, based on the radial velocities given by
Sandage \& Tammann (\cite{sandage+tammann81}) and
$H_0 = 75\,{\rm km\,s}^{-1}$.
The data from this object available up to now are rather
sparse. Schmidt et al.\ (\cite{schmidt+90}) show that the nucleus of this
galaxy is dominated by young components and suggest a recent
starburst, based on emission line spectra.
The CO molecule was observed in its lower rotational transitions
by Braine et al.\ (\cite{braine+93}) and Young et al.\ (\cite{young+95}).
These authors, however, observed just a central point or three points on an
arbitrarily chosen axis, respectively. In fact, the raster map shown by
Wielebinski et al.\ (\cite{rw+md+cn99}), which is produced from the same
data set as presented in the current paper and shown in
Fig.\ \ref{fig:n278}a,
is the first mapping of the molecular ISM in this galaxy at all.
Because of the sparse data, even basic morphological properties
are not well known.
Our data can be used to get some constraints on the orientation:
in the channel maps (not shown), the largest velocity
gradient can be found from southwest to northeast,
along a position angle of $\sim 55^{\circ}$ (ccw from north).
Hence this will be taken as position angle in the further analysis.

{\it Morphology.}
The CO(3--2) emission is distributed throughout the entire optical disk.
This is probably due to the limited spatial resolution of our data.
Two emission maxima are visible in the northwest and southeast, i.e.\ above
and below the central region along the minor axis (perpendicular to the
largest velocity gradient). The interpretation depends strongly on the
(unknown) inclination of the galaxy. In comparison with the
NIR morphology (Rhoads \cite{rhoads98}) these emission maxima most probably
arise at both ends of a molecular bar. However, more CO data at higher
angular resolution are needed for further discussion.

The intensity distribution of the CO(3--2) emission a\-long the major and
minor axis (Fig.\ \ref{fig:n278}d) is resolved and non-Gaussian.
The CO(3--2) emission is not confined to a central region, but
distributed over the whole (optically visible) galaxy. From a Gaussian
fit to the intensity distribution we would determine a deconvolved source
size of $38'' \times 39''$. When we assume a disk-like distribution, the
deconvolved source size would be rather $45'' \times 46''$, which seems
to be more appropriate for NGC\,278. This corresponds to a linear size of
$2.7\,{\rm kpc} \times 2.8\,{\rm kpc}$. It represents the extended CO(3--2)
emission, not the size of a central emission peak, as for the other objects.

{\it Intensities and Line Ratios.}
The integrated CO(3--2) line flux of NGC\,278 is
$F_{\rm CO(3-2)} = 3.7 \pm 0.3\,10^3\,{\rm Jy\,km\,s}^{-1}$.
The total power radiated in this line by is
$P_{\rm CO(3-2)} = 6.2 \pm 0.5\,10^{30}\,{\rm W}$. While this value is
similar to that for other galaxies, in NGC\,278 this power is radiated from
a much larger region.

Braine et al.\ (\cite{braine+93}) measured line intensities of 18 and
$17\,{\rm K\,km\,s}^{-1}$ for the CO(1--0) and the (2--1) line respectively.
The intensity maximum in our map is
$I_{\rm CO(3-2)} = 21 \pm 3\,{\rm K\,km\,s}^{-1}$. The spectra of Braine et
al.\ (\cite{braine+93}), however, were taken at a position which is
($-10''$, $0''$) relative to our central position which
is given in Table \ref{tab:obs}. For this position the integrated intensity
is only $15 \pm 3\,{\rm K\,km\,s}^{-1}$. Hence we estimate line ratios of
$R_{3,1} = R_{3,2} = 0.8 \pm 0.2$, which were given by Wielebinski
et al.\ (\cite{rw+md+cn99}). These line ratios are consistent with
molecular gas densities of 
$10^3 - 10^{3.5}\,{\rm cm}^{-3}$ at temperatures of $20 - 40\,{\rm K}$.

The line ratios, total flux, and the extent of the emission region
indicate that the amount of molecular gas is much smaller than in other
galaxies, but the physical conditions in the molecular gas are not too
different. This may suggest an even earlier Hubble type than Sb for this
galaxy, assuming that much of the molecular gas has already been used up
for star formation.
A more detailed investigation of the line ratios has to await a
complete mapping with higher angular resolution, including also the lower
line transitions.

{\it Kinematics.}
The lines in NGC\,278 are relatively narrow,
$\Delta V$ $\sim$ $40\,{\rm km\,s}^{-1}$.
This, as well as the low apparent rotational velocity of
$\sim 40\,{\rm km\,s}^{-1}$,
indicates that NGC\,278 is oriented relatively face-on. At a distance of
12.4\,Mpc, the angular diameter of $D_{25} = 2'\!\!.1$ corresponds to
about 8\,kpc, which is much smaller than for normal spiral galaxies,
and indicates also a lower rotational velocity. For an assumed real
rotational velocity in the order of $100\,{\rm km\,s}^{-1}$ we can
roughly estimate an inclination of about $20^{\circ} - 25^{\circ}$.

\subsection{NGC\,891}

\begin{figure*}
\resizebox{\hsize}{!}{\includegraphics*{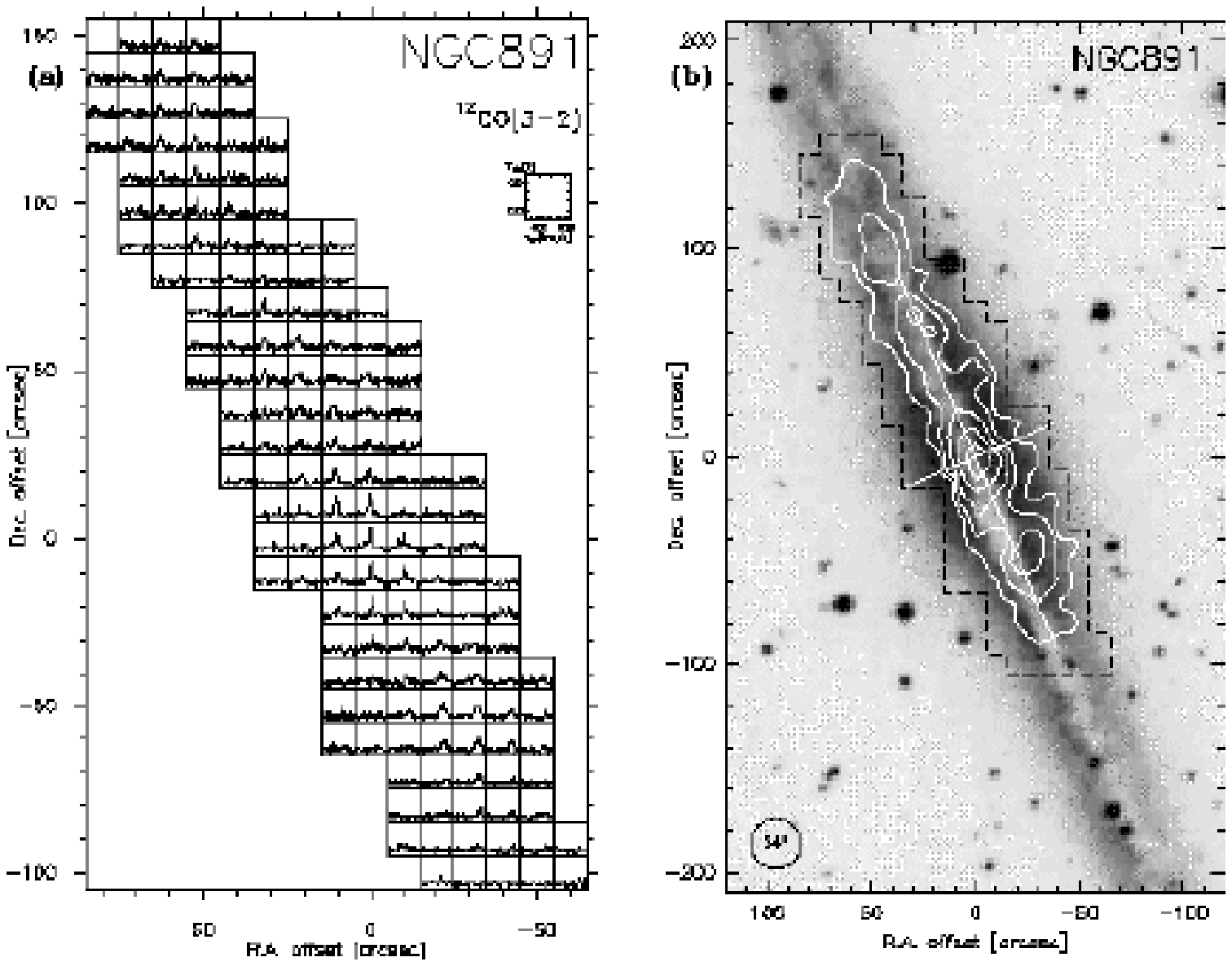}}
{\bf Fig.~3a and b.}~CO(3--2) data of NGC\,891:
{\bf a} (left) raster map of the individual spectra, with (0,0)
corresponding to the position given in Table \ref{tab:obs}.
The scale of the spectra is indicated by the small box inserted
in the upper right corner of the image.\linebreak
{\bf b} (right) contour map of the integrated intensity, overlaid on an
optical image extracted from the Digitized Sky Survey. Contour levels are
12, 24, \ldots, 60\,K\,km\,s$^{-1}$.
The region covered by the spectra shown in (a) is indicated by the dashed
polygon, and the cross shows the adopted centre and the direction
of the major and minor axis
\label{fig:n891-1}
\end{figure*}
\setcounter{figure}{2}

NGC\,891 is a nearby ($D = 9.6\,{\rm Mpc}$, consistent with
$H_0 = 75\,{\rm km\,s}^{-1}$; e.g.\ Allen et al.\ \cite{allen+78}),
nearly exactly edge-on
($i = 88^{\circ}$) Sb galaxy which is often cited as being similar to
our Milky Way. Due to this similarity and its orientation (which allows
us to study the halo gas) this object has already been investigated at
several wavelengths. Although NGC\,891 has not undergone a starburst, it
is an actively star-forming galaxy, which was not at least documented by
the explosion of SN1986J in its southern half (e.g.\ Ball \& Kirk
\cite{ball+kirk95}).
Garc\'{\i}a-Burillo et al.\ (\cite{garcia+92}) and
Garc\'{\i}a-Burillo \& Gu\'elin (\cite{garcia+guelin95}) have mapped the
CO(2--1) and the CO(1--0) lines in this galaxy using the 30\,m telescope
on Pico Veleta. Interferometric observations were also published by
Scoville et al.\ (\cite{scoville+93})

{\it Morphology.}
In
\setcounter{figure}{3}
Fig.\ \arabic{figure}
\setcounter{figure}{2}
we show the first published CO(3--2) data
of this prototype of a non-interacting, Milky Way-like
spiral. The CO(3--2) emission seems to be relatively weak compared to
the data on starburst galaxies already published. The peak temperatures do
not exceed $T_{\rm mb} \sim 0.7\,{\rm K}$. In the central region, the
(relatively narrow) main emission component is accompanied by a second,
very weak and broad line, which represents the velocity components due
to the central bar and which were already investigated for the lower
CO transitions by Garc\'{\i}a-Burillo \& Gu\'elin (\cite{garcia+guelin95}).
While the CO(1--0) emission is more concentrated in the inner disk and shows
a pronounced central peak, there is an interesting morphological similarity
between the CO(2--1) and the CO(3--2) emission:
intense, but not very extended emission regions in the centre and the
southern half (at a galactocentric radius of about $50''$, and a less intense,
but more extended emission region in the northern half where CO(3--2) emission
is detectable at radii larger than $2'$. We note that our observations did
not reach the edge of the emission at $160''$ from the centre. The shift
of the southern emission peak off the major axis in western direction is
visible in both the CO(2--1) and the CO(3--2) maps.

In order to estimate the size and the line flux of the central peak in
NGC\,891, we had to subtract the disk emission which can be found on the
same line-of-sight in the fore- and background. Therefore we constructed
a radial model of the off-centre emission, based on the measured
intensities at $|x| > 30''$, and used this model to estimate the underlying
disk emission at small radii. After subtraction of this emission from the
major axis intensity distribution, we find a deconvolved extent for
the central CO(3--2) peak of $17'' \times 8''$. This corresponds to
$800\,{\rm pc} \times 360\,{\rm pc}$ at the distance of NGC\,891, with
the larger value along the major axis.
These values concern only the central peak -- as noted above, emission
can be detected along the whole (observed) major axis.

\begin{figure*}
\resizebox{\hsize}{!}{\includegraphics*{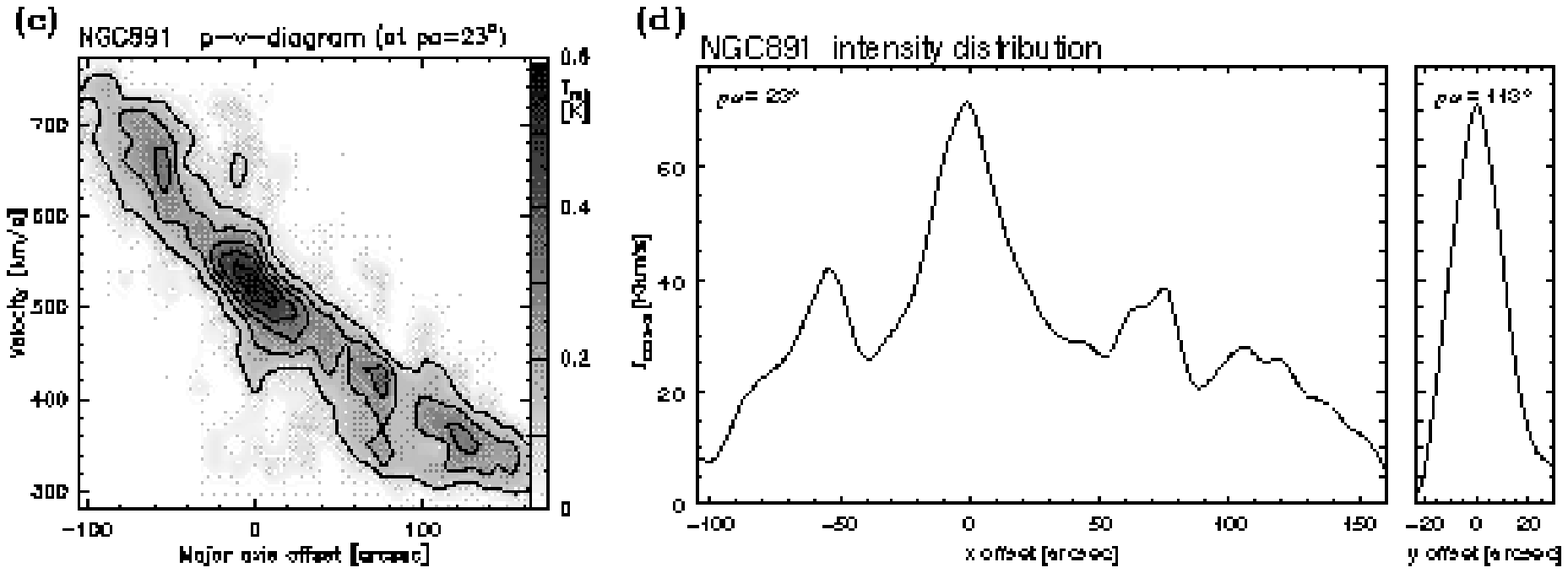}}
{\bf Fig.~3c and d.}~CO(3--2) data of NGC\,891:
{\bf c} (left) position-velocity diagram along the major axis. Contours
are 0.1, 0.2, \ldots 0.5\,K.
{\bf d} (right) intensity distribution along the major and minor
axis, respectively
\label{fig:n891-2}
\end{figure*}
\setcounter{figure}{3}

{\it Intensities.}
Integrated over the whole region covered by our CO(3--2) mapping,
we determine a total line flux of
$F_{\rm CO(3-2)}^{\rm tot} = 2.5 \pm 0.2\,10^4\,{\rm Jy\,km\,s}^{-1}$.
This corresponds to a total power emitted in the CO(3--2) line of
$P_{\rm CO(3-2)}^{\rm tot} = 2.5 \pm 0.2\,10^{31}\,{\rm W}$ at the assumed
distance of $9.6\,{\rm Mpc}$.
Since the emission does probably extend to even larger radii along the
major axis, this is a lower limit. However, the amount of
missed CO flux is smaller than a few percent. For a comparison
of the integrated properties of this galaxy with the other objects in the
sample, we estimate also the line flux of the central concentration and
get, after subtraction of the disk emission in the fore- and background,
$F_{\rm CO(3-2)}^{\rm cen} \sim 4.7 \pm 0.6\,10^3\,{\rm Jy\,km\,s}^{-1}$,
hence a total power radiated by the central regions of
$P_{\rm CO(3-2)}^{\rm cen} \sim 4.7 \pm 0.6\,10^{30}\,{\rm W}$.

{\it Line Ratios.}
The CO data for lower transitions do not show very high
brightness temperatures: these reach $T_{\rm mb} \sim 1.4\,{\rm K}$ for the
CO(2--1) line and $T_{\rm mb} \sim 1.3\,{\rm K}$ for the CO(1--0) line
(Garc\'{\i}a-Burillo \& Gu\'elin \cite{garcia+guelin95}).
Since the higher value for the CO(2--1) line is due to the
smaller beam and the therefore smaller beam dilution factor, the
temperatures give a line ratio of $R_{3,1} \sim 0.5$.

In the centre of NGC\,891, Garc\'{\i}a-Burillo et
al.\ (\cite{garcia+92}) measured an integrated CO intensity for the
(1--0) line of about $180\,{\rm K\,km\,s}^{-1}$ (in the $T_{\rm mb}$-scale).
With a similar telescope beam size we find
$I_{\rm CO(3-2)} = 71 \pm 9\,{\rm K\,km\,s}^{-1}$,
which yields a line ratio of only $R_{3,1} = 0.4 \pm 0.1$. This is smaller
than the value for $R_{3,1}$ we would obtain from the peak temperatures,
which means that the broad low-intensity spectral component from the
central bar and/or central molecular disk is weaker in the CO(3--2) line.
This indicates lower excitation temperatures in this compenent than
in the very centre. Furthermore, this line ratio is also much
lower than that for the other galaxies in our sample, which typically
show line ratios of $\sim 1$ or higher.
In the disk, this ratio seems to increase slightly up to
$R_{3,1} = 0.5 \pm 0.1$; the difference to the central region, however,
is not significant. From these low line ratios we estimate ${\rm H}_2$
densities $< 10^3\,{\rm cm}^{-3}$ for the disk as well as for
the central region of NGC\,891.

{\it Kinematics.}
The position-velocity diagram
(Fig.\ \arabic{figure}c)
does not show the steep velocity gradient of the central bar or molecular
disk (as seen by Garc\'{\i}a-Burillo et al.\ \cite{garcia+92} for the lower
transitions). The kinematics of the CO(3--2) rather follow that of the
\ion{H}{i} emission (e.g.\ Rupen \cite{rupen91}). This again suggests that
highly excited molecular gas, while it is ubiquitous in the spiral arms,
plays only a minor role in the central molecular gas disk or in the
molecular bar of NGC\,891. This is different from other galaxies,
where most of the CO(3--2) is emitted from the central region, as
opposed to the spiral arms.

The line width is $\Delta\,V \sim 100\,{\rm km\,s}^{-1}$ at the central
position, and decreases to the north and the south to about
80\,km\,s$^{-1}$. This concerns only the narrow main component. As already
mentioned, the second, weaker, component due to noncircular velocities in
the central region is much broader.

\subsection{Maffei\,2}

\begin{figure*}
\resizebox{\hsize}{!}{\includegraphics*{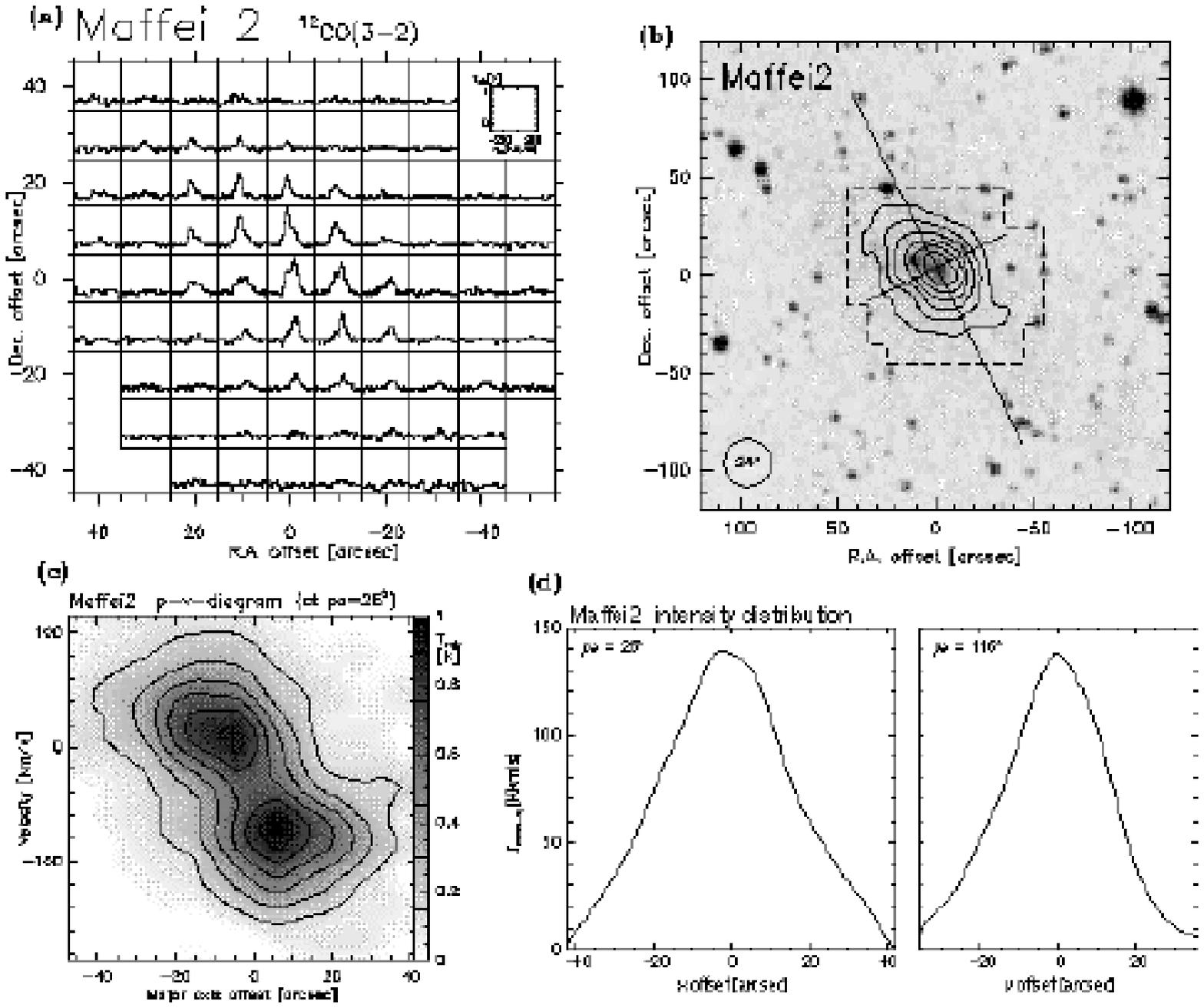}}
\caption{CO(3--2) data of Maffei\,2:
{\bf a} (upper left) raster map of the individual spectra, with (0,0)
corresponding to the position given in Table \ref{tab:obs}.
The scale of the spectra is indicated by the small box inserted
in the upper right corner of the image.
{\bf b} (upper right) contour map of the integrated intensity, overlaid on an
optical image extracted from the Digitized Sky Survey. Contour levels are
20, 40, \ldots, 120\,K\,km\,s$^{-1}$.
The region covered by the spectra shown in (a) is indicated by the dashed
polygon, and the cross shows the reference position for data analysis
(see text) and the direction of the major and minor axis.
{\bf c} (lower left) position-velocity diagram along the major axis. Contours
are 0.15, 0.3, \ldots 0.9\,K.
{\bf d} (lower right) intensity distribution along the major and minor
axis, respectively.
Note that for figure parts {\bf c} and {\bf d} the zero position
differs from the position (0,0) in figure parts {\bf a} and {\bf b} by
$(\Delta {\rm R.A.}, \Delta{\rm Dec.}) = (-0''\!\!.8, 4''\!\!.0)$
}
\label{fig:maf2}
\end{figure*}

Maffei\,2 is a nearby starburst galaxy, classified as SAB(rs)bc, and
located near the galactic plane ($b = -0^{\circ}\!\!.5$;
e.g.\ Spinrad et al.\ \cite{spinrad+71}). As distance to this object we
assume the commonly used value of $D = 5\,{\rm Mpc}$ to be consistent
with previous publications (e.g.\ Hurt et al.\ \cite{hurt+93};
H\"uttemeister et al.\ \cite{huette+95}).
Due to its location in the sky and its systemic velocity
($v_{\rm lsr} = -36\,{\rm km\,s}^{-1}$), molecular line observations of
this object are rare. Published CO(1--0) and (2--1) data exist from Rickard
et al. (\cite{rickard+77}), Sargent et al.\ (\cite{sargent+85}), and
Weliachew et al.\ (\cite{weliachew+88}). The CO(3--2) line was observed by
Hurt et al.\ (\cite{hurt+93}) with the CSO. The CO(4--3) line was also
observed in the meantime, and will be published together with more recent
CO(1--0) and (2--1) data (Walsh et al., in prep.).

{\it Morphology.}
Our map (Fig.\ \ref{fig:maf2}a) covers a region which is about twice as
large as the region mapped by Hurt et al.\ (\cite{hurt+93}). Our data
confirm that the CO(3--2) emission is -- as the lower transitions --
concentrated to the inner disk of this galaxy. The peak brightness
temperatures are between 1 and 1.5\,K, similar to the values of
Hurt et al.\ (\cite{hurt+93}) when transformed to the $T_{\rm mb}$-scale.
Their intensities, however, are slightly higher than our values.
For the central position, which is also the maximum of our map, we measure
a value of $I_{\rm CO(3-2)} = 139 \pm 10\,{\rm K\,km\,s}^{-1}$.
The highest value
of Hurt et al.\ (\cite{hurt+93}) exceeds $170\,{\rm K\,km\,s}^{-1}$
(after smoothing to $24''$) and is therefore about 20\,\% higher.
These differences may have several causes -- pointing or calibration
uncertainties in either of both data sets. Hurt et al.\ (\cite{hurt+93})
do not give efficiency values, and their Fig.\ 1 indicates pointing
difficulties. For Maffei\,2, all data presented here were taken after
the encoder problem mentioned in Sect.\ \ref{section:obs} has been
corrected, hence the pointing accuracy for the Maffei2 data set was $3''$. 

The reference position of our observations which is listed in
Table \ref{tab:obs} was chosen at the coordinates of Weliachew et
al.\ (\cite{weliachew+88}). For the data analysis
(Figs.\ \ref{fig:maf2}c and d) we shifted the reference point slightly
in northeastern direction onto the CO(3--2) maximum, located at
$(\Delta {\rm R.A.}, \Delta{\rm Dec.}) = (-0''\!\!.8, 4''\!\!.0)$
relative to the (0,0) position. This position is identical with the
reference position used by Hurt et al.\ (\cite{hurt+93}).

From Gaussian fits we estimate a deconvolved source size of
$32'' \times 21''$, corresponding to a linear size of
$780\,{\rm pc} \times 520\,{\rm pc}$ at the distance of Maffei\,2.
Even if the intensity distribution (Fig.\ \ref{fig:maf2}b) is centrally
peaked, the shape of the individual spectra as well as the position-velocity
diagram along the major axis suggest a ring-like distribution of
the highly excited molecular gas, with a diameter of $10'' - 20''$.
This is too small to be resolved by our telescope beam.
A two-peaked appearance of the position-velocity diagram along the minor
axis (not shown) suggests a more complicated kinematical structure, perhaps
a bar which is inclined with respect to the main disk. However,
the bar-like structure as seen in CO(1--0) interferometric maps
(Hurt \& Turner \cite{hurt+turner91}) is not present in our data.

{\it Intensities and Line Ratios.}
The integrated line flux of Maffei\,2 in the CO(3--2) line is
$F_{\rm CO(3-2)} = 1.87 \pm 0.08\,10^4\,{\rm Jy\,km\,s}^{-1}$,
which corresponds at the distance to Maffei\,2 to a total power of
$P = 5.1 \pm 0.3\,10^{30}\,{\rm W}$.
This value for $P_{\rm CO(3-2)}$ is similar to most other galaxies in
our sample.

From the CO(1--0) data of Weliachew et al.\ (\cite{weliachew+88}) and our
observations we have determined a line ratio of $R_{3,1} = 1.3 \pm 0.2$ in
the centre of Maffei\,2 and $R_{3,1} = 0.8 \pm 0.2$ in the disk (at radii of
$\sim 30''$). This suggests high excitation temperatures in the starburst
region. From a simple LVG analysis we derive densities of
$n({\rm H}_2) \sim 10^{3.5}\,{\rm cm}^{-3}$ in the centre, falling of rapidly
to $n({\rm H}_2) < 10^3\,{\rm cm}^{-3}$. The kinetic temperatures in the
centre must be $T > 50\,{\rm K}$.
Interestingly the estimated line ratios for Maffei\,2 are higher than those
for other galaxies with a ring-like molecular gas distribution, and resemble
those for objects showing a more centrally peaked distribution.

{\it Kinematics.}
The spectra from the central region show linewidths of
$\Delta\,V \sim 150\,{\rm km\,s}^{-1}$. This can be attributed to two
individual components, with line parameters of
$V_0 \sim -85\,{\rm km\,s}^{-1}$ and $\Delta\,V \sim 70\,{\rm km\,s}^{-1}$
for the northeastern and
$V_0 \sim 10\,{\rm km\,s}^{-1}$ and $\Delta\,V \sim 90\,{\rm km\,s}^{-1}$
for the southwestern component.

\subsection{IC\,342}

\begin{figure*}
\resizebox{\hsize}{!}{\includegraphics*{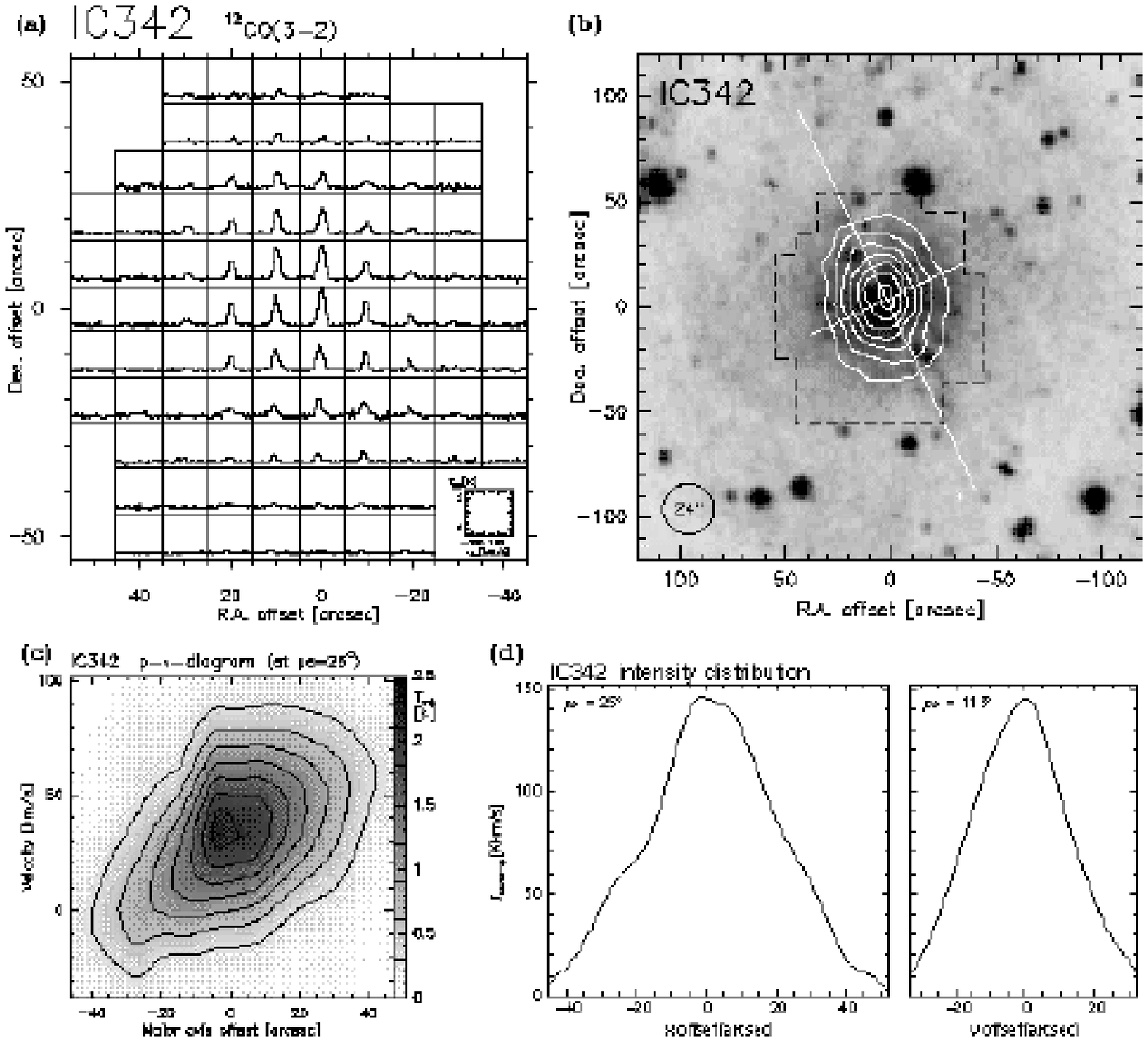}}
\caption{CO(3--2) data of IC\,342:
{\bf a} (upper left) raster map of the individual spectra, with (0,0)
corresponding to the position given in Table \ref{tab:obs}.
The scale of the spectra is indicated by the small box inserted
in the lower right corner of the image.
{\bf b} (upper right) contour map of the integrated intensity, overlaid on an
optical image extracted from the Digitized Sky Survey. Contour levels are
20, 40, 60, \ldots, 140\,K\,km\,s$^{-1}$.
The region covered by the spectra shown in (a) is indicated by the dashed
polygon, and the cross shows the reference position for data analysis
(see text) and the direction of the major and minor axis.
{\bf c} (lower left) position-velocity diagram along the major axis. Contours
are 0.3, 0.6, \ldots 2.1\,K.
{\bf d} (lower right) intensity distribution along the major and minor
axis, respectively.
Note that for figure parts {\bf c} and {\bf d} the zero position
differs from the position (0,0) in figure parts {\bf a} and {\bf b} by
$(\Delta {\rm R.A.}, \Delta{\rm Dec.}) = (1''\!\!.7, 3''\!\!.6)$
}
\label{fig:ic342}
\end{figure*}

The SBcd galaxy IC\,342 is one of the nearest grand-design spiral galaxies.
Distance estimates range from 1.8\,Mpc (McCall \cite{mccall89}) -- this
value implies that infrared and CO luminosities are similar to those of
the Milky Way -- to 8\,Mpc (Sandage \& Tammann \cite{sandage+tammann74}).
In order to be
consistent with previous molecular line studies, we assume
a distance of $D \sim 4.5\,{\rm Mpc}$ (Xie et al.\ \cite{xie+94};
Steppe et al.\ \cite{steppe+90}, and references therein).

IC\,342 is oriented almost face-on ($i=25^{\circ}$) and exhibits a
nuclear starburst. Young \& Scoville (\cite{young+scoville82}) observed the
CO(1--0) transition along two perpendicular strips in this galaxy.
More recent interferometric observations were performed by
Ishizuki et al.\ (\cite{ishizuki+90}).
Xie et al.\ (\cite{xie+94}) made a study
of the two lowest rotational transitions of \element[][12]{CO} and
\element[][13]{CO} in NGC\,2146 and IC\,342. Also the CO(3--2) line
(Steppe et al.\ \cite{steppe+90}), and even CO(4--3) emission
(G\"usten et al.\ \cite{guesten+93}), was already observed. Recently,
Schulz et al.\ (\cite{schulz+01}) observed \element[][13]{CO} in the
(2--1) and (3--2) transitions using the IRAM 30-m telescope.

{\it Morphology.}
We mapped IC\,342 out to a galactocentric distance of about 1.3\,kpc.
Within this mapped region, our data show a concentration of the
CO(3--2) emission to the inner disk of IC\,342.
The emitting region is slightly elongated along the major axis
of the galaxy, with a deconvolved source size of $38'' \times 24''$,
corresponding to $840\,{\rm pc} \times 520\,{\rm pc}$ at the assumed
distance of 4.5\,Mpc. In CO data with higher spatial resolution
(e.g.\ Schulz et al., in prep.), this elongated structure consists of at
least two emission maxima, separated by $\sim 8''$. The southwest of these
maxima corresponds to our original reference position (see
Table \ref{tab:obs}).
For the data analysis (Figs.\ \ref{fig:ic342}c and d) we shifted the
reference point by $4''$ along the major axis in northeastern direction,
between these two maxima. This position, at
$(\Delta {\rm R.A.}, \Delta{\rm Dec.}) = (1''\!\!.7, 3''\!\!.6)$,
corresponds to the CO(3--2) maximum when observed with the angular
resolution of the HHT.

{\it Intensities and Line Ratios.}
We find a peak intensity of the CO(3--2) line of
$I_{\rm CO(3-2)} = 148 \pm 6\,{\rm K\,km\,s}^{-1}$.
Steppe et al.\ (\cite{steppe+90})
obtained higher values (with their smaller beam) especially in the
northeastern part of our map. This difference may reflect the beam dilution
resulting from the barred distribution of the molecular gas, as seen
in the CO(1--0) line by Ishizuki et al.\ (\cite{ishizuki+90}).
The CO(2--1) observations of Xie et al.\ (\cite{xie+94}) which are
obtained with a similar beam size as our CO(3--2) data allow us to directly
estimate line ratios. We find $R_{3,2} = 1.2 \pm 0.1$ in the centre, but values
slightly smaller than 1 at $30''$ distance from the centre along the major
axis. For the (2--1)/(1--0) ratio Xie et al.\ (\cite{xie+94}) give
$R_{2,1} = 1.1 \pm 0.3$ for data smoothed to $45''$ HPBW, which corresponds
to $R_{3,1} \sim 1.3$ and $\sim 1$ in the centre and the disk respectively.
These values suggest a kinetic temperature of $T > 50\,{\rm K}$ and gas
densities of $n({\rm H}_2) \sim 10^{3.5}\,{\rm cm}^{-3}$ in the centre,
but lower values in the disk. Thus the physical conditions in the
molecular gas are comparable to other objects such as Maffei\,2.
We find an integrated line flux of
$F_{\rm CO(3-2)} = 2.19 \pm 0.06\,10^4\,{\rm Jy\,km\,s}^{-1}$.
With the assumed distance of 4.5\,Mpc to IC\,342, this translates to a
total power emitted in the CO(3--2) line of
$P_{\rm CO(3-2)} = 4.9 \pm 0.2\,10^{30}\,{\rm W}$.
This value is similar to those for other spiral galaxies in our sample,
but note the distance uncertainties mentioned at the beginning of this
subsection (the distance enters as $D^2$ into the total power).
The ratio $P_{\rm CO(3-2)} / (l_{\rm x} \times l_{\rm y})$, i.e.\ the total
power divided by the size of the emitting region, is
$11.2\,10^{30}\,{\rm W\,kpc}^{-2}$, which is also of the same
order as for some other objects in our sample.

{\it Kinematics.}
A velocity gradient, consistent with normal galactic rotation, is
visible along the major axis (Fig.\ \ref{fig:ic342}c).
Due to the small inclination ($i \sim 25^{\circ}$) the line width is
small, especially when compared to the edge-on objects in our sample.
We determined it by a Gaussian fit to $\Delta\,V \sim 60\,{\rm km\,s}^{-1}$.
It is relatively constant over the region where we detected CO(3--2) emission.
Since IC\,342 is classified as a barred spiral, this small line width
indicates either a very particular orientation of the bar, or the absence
of highly excited CO in the bar, since otherwise the various possible orbits
would lead to a higher velocity dispersion near the centre of this galaxy.

\subsection{NGC\,2146}

\begin{figure*}
\resizebox{\hsize}{!}{\includegraphics*{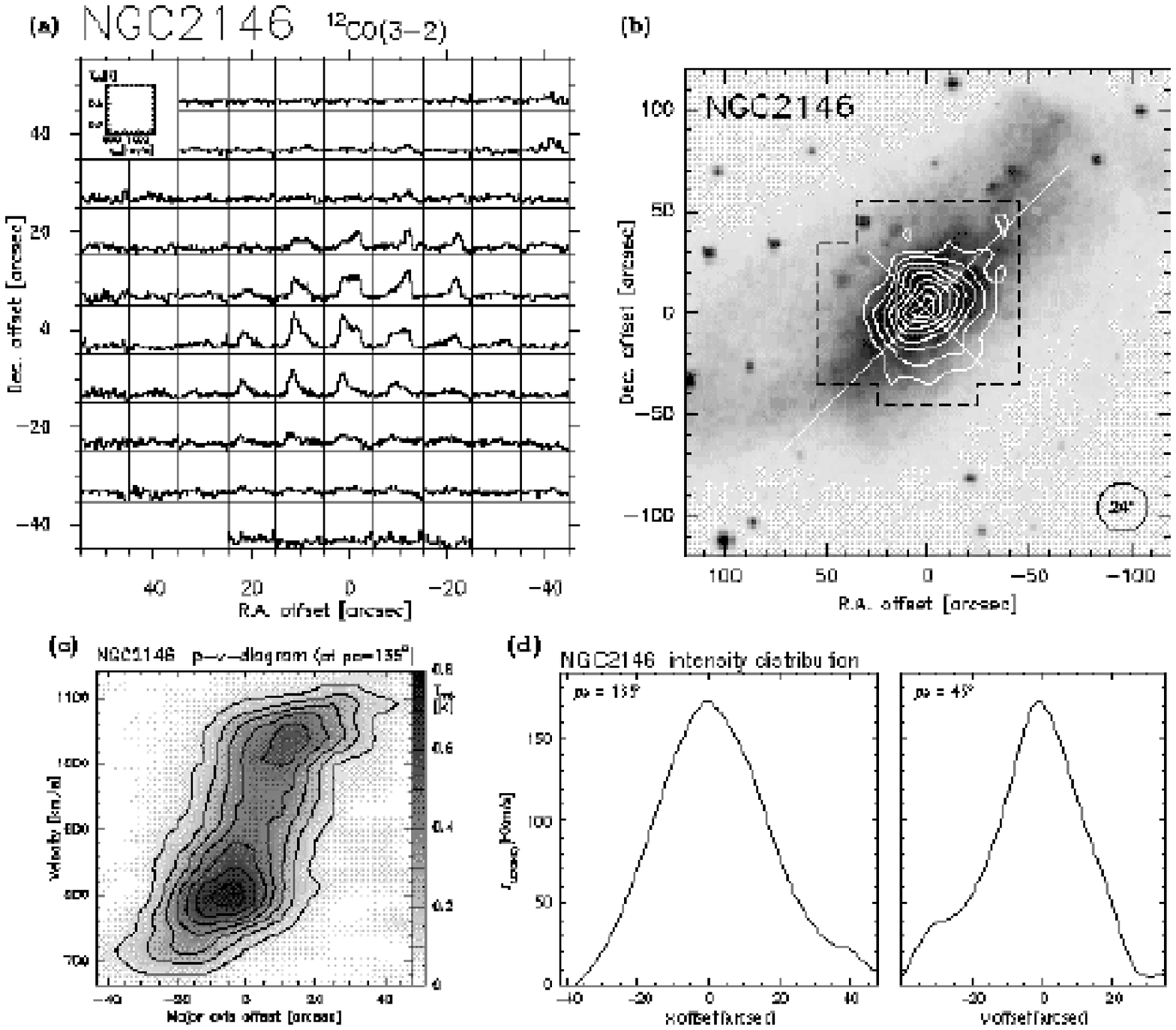}}
\caption{CO(3--2) data of NGC\,2146:
{\bf a} (upper left) raster map of the individual spectra, with (0,0)
corresponding to the position given in Table \ref{tab:obs}.
The scale of the spectra is indicated by the small box inserted
in the upper left corner of the image.
{\bf b} (upper right) contour map of the integrated intensity, overlaid on an
optical image extracted from the Digitized Sky Survey. Contour levels are
20, 40, 60, \ldots, 160\,K\,km\,s$^{-1}$.
The region covered by the spectra shown in (a) is indicated by the dashed
polygon, and the cross shows the reference position for data analysis
(see text) and the direction of the major and minor axis.
{\bf c} (lower left) position-velocity diagram along the major axis. Contours
are 0.1, 0.2, \ldots 0.7\,K.
{\bf d} (lower right) intensity distribution along the major and minor
axis, respectively.
Note that for figure parts {\bf c} and {\bf d} the zero position differs
from the position (0,0) in figure parts {\bf a} and {\bf b} by
$(\Delta {\rm R.A.}, \Delta{\rm Dec.}) = (1''\!\!.5, 2''\!\!.0)$
}
\label{fig:n2146}
\end{figure*}

NGC\,2146 is the most distant object in our sample. Distances cited in
the literature range from 14 to 21\,Mpc; we adopt 17\,Mpc throughout this
study. This galaxy is classified as SB(s)ab\,pec, and with an inclination
of $65^{\circ}$ it still can be considered as oriented edge-on.
NGC\,2146 is a starburst galaxy, showing a superwind driven by violent
star formation in its central region (Armus et al.\ \cite{armus+95}),
but has no obvious companion --- a fact that distinguishes NGC\,2146
from other known nearby starburst galaxies. Despite its remarkable
appearance in the optical, the molecular line data on this galaxy is sparse.
Xie et al.\ (\cite{xie+94}) observed the two lowest rotational transitions
of \element[][12]{CO} and \element[][13]{CO}, but only along the major and
minor axis.
A few years earlier, Jackson \& Ho (\cite{jackson+ho88}) obtained an
interferometer map of the (1--0) transition. Recently further CO(1--0) and
(2--1) maps, covering the whole optical disk, were obtained at the
IRAM 30-m telescope (Dumke et al., in prep.).
There are also interferometric data for the central
region made with the Plateau de Bure interferometer (Greve et
al.\ \cite{greve+00}; Greve et al.\ in prep.).
Here we present the first CO(3--2) data of this galaxy.

{\it Morphology and Intensities.}
Because of some pointing problems during the first observing run in April 1998
(see Sect.\ \ref{section:obs}) some spectra in the northern half of the mapped
region had to be discarded during the data reduction, but were re-observed.
The similarity between the line shapes of our final CO(3--2) map and
the CO(2--1) data of Xie et al.\ (\cite{xie+94}) shows that the pointing
accuracy finally reached is relatively good.

\begin{figure*}
\resizebox{12cm}{!}{\includegraphics*{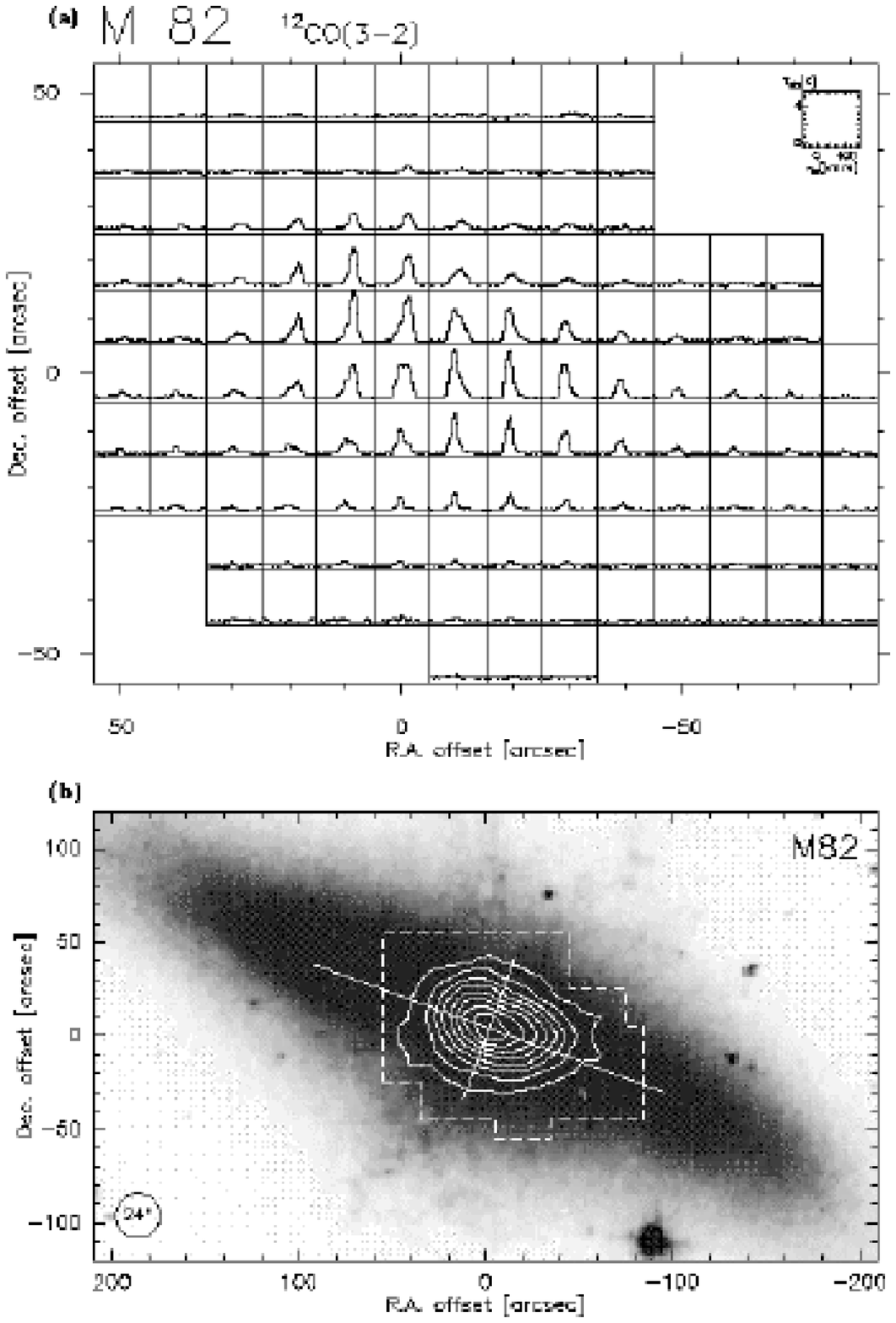}}
\hfill
\parbox[b]{55mm}{
{\bf Fig.~7a and b.}~CO(3--2) data of M\,82:
{\bf a} (upper) raster map of the individual spectra, with (0,0)
corresponding to the position given in Table \ref{tab:obs}.
The scale of the spectra is indicated by the small box inserted
in the upper right corner of the image.
{\bf b} (lower) contour map of the integrated intensity, overlaid on an
optical image extracted from the Digitized Sky Survey. Contour levels are
70, 140, 220, 300, \ldots, 700\,K\,km\,s$^{-1}$.
The region covered by the spectra shown in (a) is indicated by the dashed
polygon, and the cross shows the reference position for data analysis
and the direction of the major and minor axis
\label{fig:m82-1}
}
\end{figure*}
\setcounter{figure}{6}

The reference position for our observations (Table \ref{tab:obs}) was that
given by Young et al.\ (\cite{young+95}). For the data analysis (see
Figs.\ \arabic{figure}c and d) we shifted the reference point by
$(\Delta {\rm R.A.}, \Delta{\rm Dec.}) = (1''\!\!.5, 2''\!\!.0)$ onto the
measured CO(3--2) maximum, and the position angle was assumed to be
$135^{\circ}$. While the optical position angle is $143^{\circ}$, the
major axis of the CO(1--0) distribution as found by interferometric
observations (Jackson \& Ho \cite{jackson+ho88}) is closer to $135^{\circ}$,
and Xie et al.\ (\cite{xie+94}) used even $128^{\circ}$ for their observations.

The CO(3--2) emission of NGC\,2146 is elongated along the major axis of
the galaxy, with a deconvolved source size of $29'' \times 17''$. This
corresponds to $2.4\,{\rm kpc} \times 1.4\,{\rm kpc}$ at a distance of
17\,Mpc. The source extent along the minor axis is rather large. However,
if we take the orientation ($i = 65^{\circ}$) and the peculiar optical
appearance of this galaxy into account, it seems unlikely that the emission
should be confined to a line along the position angle.
The integrated line flux is $1.94 \pm 0.14\,10^4\,{\rm Jy\,km\,s}^{-1}$,
but the large distance to NGC\,2146 makes this the object with the largest
power emitted in the CO(3--2) line. With
$P_{\rm CO(3-2)} = 6.1 \pm 0.5\,10^{31}{\rm W}$ this galaxy is about four
times as luminous as the prototype starburst galaxy M\,82.
This points to a strongly enhanced starburst activity in this object.
This activity may probably be caused by a merging event, which can account
for the disturbed appearance. It may also be the cause
for physical conditions in the molecular gas that allow emission from
the $J = 3$ level of CO over a larger area. Further,
such a merging event may also have caused a gas outflow perpendicular to the
major axis of the galaxy which may also be the reason for the large measured
source size.

\begin{figure*}
\resizebox{\hsize}{!}{\includegraphics*{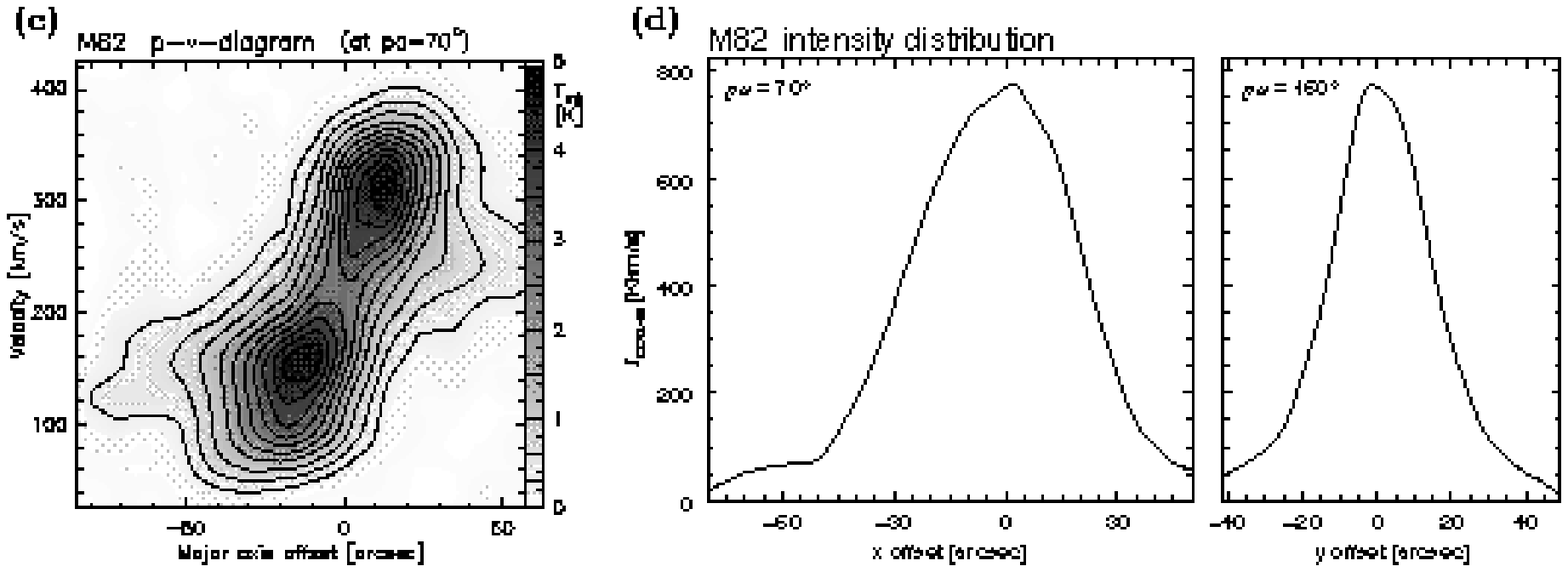}}
{\bf Fig.~7c and d.}~CO(3--2) data of M\,82:
{\bf c} (left) position-velocity diagram along the major axis. Contours
are 0.3, 0.6, 1.0, 1.5, \ldots 4.5\,K.
{\bf d} (right) intensity distribution along the major and minor
axis, respectively. Note that for figure parts {\bf c} and {\bf d} the
zero position differs from the position (0,0) in figure parts {\bf a} and
{\bf b} by $(\Delta {\rm R.A.}, \Delta{\rm Dec.}) = (-2''\!\!.0, 4''\!\!.0)$
\label{fig:m82-2}
\end{figure*}
\setcounter{figure}{7}

When we compare the total power emitted in the CO(3--2) line with the
size of the region in which it is produced, we find that the ratio of these
two numbers is similar to the values found in the central regions of
Maffei\,2 and IC\,342. Hence NGC\,2146 is capable of maintaining physical
gas conditions which are typical for starbursts over a large region,
with a diameter of more than 2\,kpc and a thickness of more than 1\,kpc.

{\it Line Ratios.}
The velocity-integrated CO(3-2) intensity in the centre is
$I_{\rm CO(3-2)} = 172 \pm 17\,{\rm K\,km\,s}^{-1}$. For the CO(2--1)
transition Xie et al.\ (\cite{xie+94}) have measured
$I_{\rm CO(2-1)} = 123.3 \pm 27.4 \,{\rm K\,km\,s}^{-1}$ at a similar
angular resolution. Since their relatively large error includes also
calibration uncertainties, it will only partly enter into our line ratio
calculation, which yields a value of $R_{3,2} = 1.4 \pm 0.2$ in the
centre, and $1.0 \pm 0.2$ at a $30''$ offset along the major axis.
For the (2--1)/(1--0) line ratio Xie et al.\ give $R_{2,1} = 0.95 \pm 0.23$
in the centre (for a resolution of $45''$ HPBW).
From these values we derive a value for the (3--2)/(1--0) line ratio
in the centre of this galaxy of $R_{3,1} = 1.3 \pm 0.2$.
This line ratio fits well with those of the other starburst
galaxies in our sample, indicating similar densities and temperatures.
Mauersberger et al.\ (\cite{mauersberger+99}) measured a slightly higher
value for $I_{\rm CO(3-2)}$ and derived therefore a line ratio of
$I_{3,1} \sim 1.6$.

{\it Kinematics.}
The kinematics of the highly excited CO gas in NGC 2146 show rigid rotation
with a large velocity gradient in the inner disk, reaching rotation
velocities of $\pm 150\,{\rm km\,s}^{-1}$ at a radius of $15''$. The
position-velocity diagram along the minor axis, which is not shown here,
shows further a large line width in the centre, indicating noncircular
velocities. A fit to the central spectrum yields
$\Delta\,V \sim 280\,{\rm km\,s}^{-1}$ and hence the largest line-width
among the galaxies in our sample. The line-shape, however, is clearly
non-Gaussian. A two-component fit yields a line-width of about
130\,km\,s$^{-1}$ for the component that is more prominent in the
southeastern half (with $V_0 \sim 820\,{\rm km\,s}^{-1}$) and
150\,km\,s$^{-1}$ for the northwestern half (with
$V_0 \sim 990\,{\rm km\,s}^{-1}$).

All this points to NGC\,2146 as a galaxy with violent star formation
over a large fraction of its inner disk. Since this galaxy has no
visible companion, the most reasonable explanation is a recent merger,
which should be modelled with the help of interferometric
line observations of \ion{H}{i} and CO.

\subsection{M\,82}

The prototypical nearby starburst galaxy M\,82 is classified as I0 and
has an inclination of $81^{\circ}$. We assume a distance of 3.2\,Mpc,
consistent with the value given by Tammann \& Sandage
(\cite{tammann+sandage68}).
It is one of the most frequently observed extragalactic sources, because
of its relatively high intensity in practically all wavelength ranges.
Different transitions of CO were observed
and analysed by several authors. Knapp et al.\ (\cite{knapp+80})
noted an unusual line ratio of $R_{2,1} \sim 2$ in the inner part of this
galaxy (which has not been confirmed by more recent observations).
Wild et al.\ (\cite{wild+90}), Turner et al.\ (\cite{turner+90}),
and Lo et al.\ (\cite{lo+90}) observed the CO(3--2) emission at selected
positions in M\,82. A CO(3--2) map of the inner disk ($r \le 30''$) was
presented by Tilanus et al.\ (\cite{tilanus+91}). The (4--3) transition
has been observed by G\"usten et al.\ (\cite{guesten+93}) and
White et al.\ (\cite{white+94}). Recently the HHT was used for
CO(7--6) observations of this object (Mao et al.\ \cite{mao+00}).

{\it Morphology.}
In this paper we present the first CO(3--2) map of M\,82 which covers
the object out to a radius of more than $1'$ ($> 900\,{\rm pc}$).
The CO(3--2) emission is
strongly enhanced in the vicinity of the centre, but it is still detected
at a high level ($I_{\rm CO(3-2)} = 30 - 50\,{\rm K\,km\,s}^{-1}$) out to
the edges of the observed region. The double-peak structure seen in
$^{13}$CO(3--2) data and in maps of other transitions
(Mao et al.\ \cite{mao+00}) is not
visible in the total intensity map of the $^{12}$CO(3--2) emission
(Fig.\ \arabic{figure}b),
but in the position-velocity diagram
(Fig.\ \arabic{figure}c).
This points to differences in the excitation
conditions and/or optical depths in the molecular lines between the
centre and the molecular ring structure seen in the two peaks.

\begin{figure*}
\resizebox{12cm}{!}{\includegraphics*{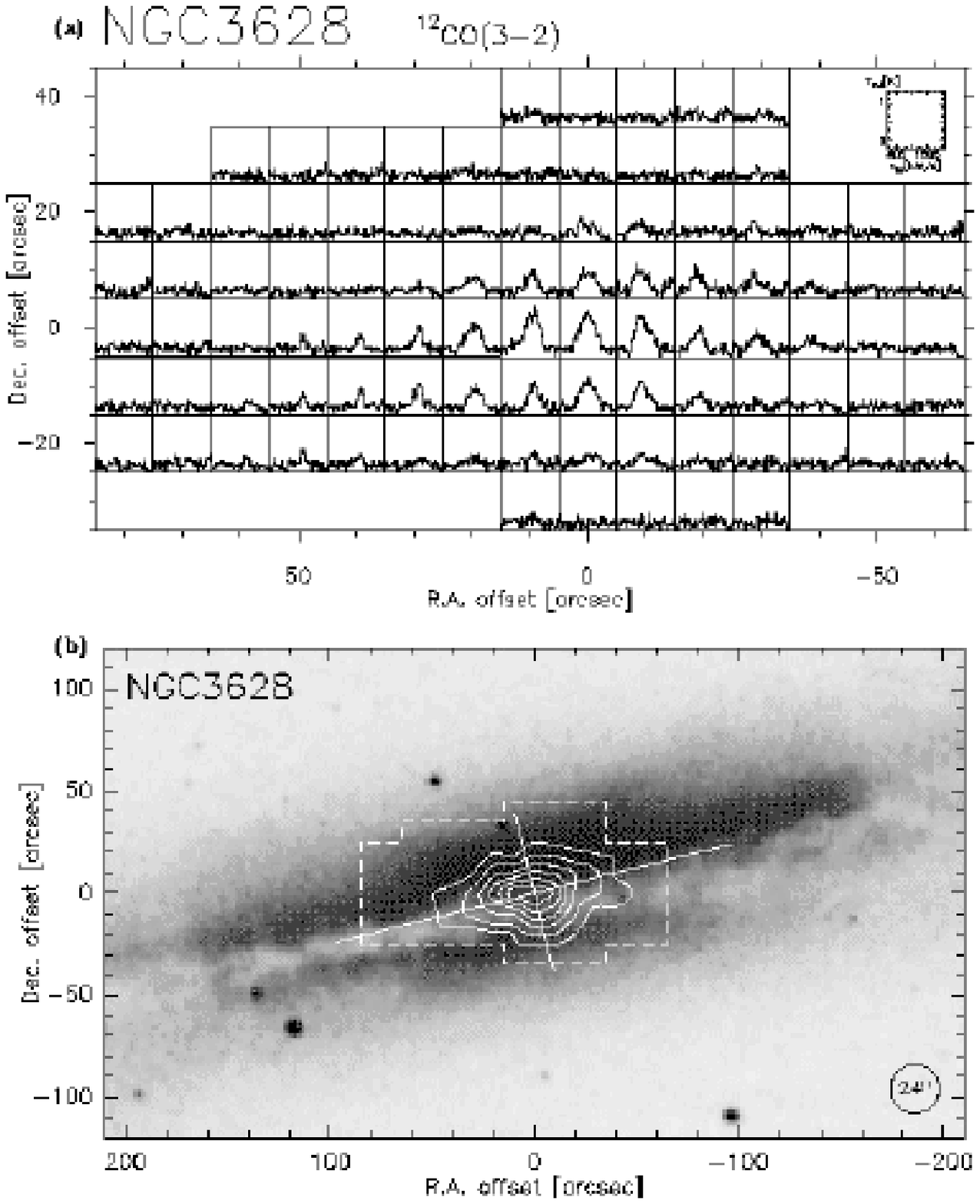}}
\hfill
\parbox[b]{55mm}{
{\bf Fig.~8a and b.}~CO(3--2) data of NGC\,3628:
{\bf a} (upper) raster map of the individual spectra, with (0,0)
corresponding to the position given in Table \ref{tab:obs}.
The scale of the spectra is indicated by the small box inserted
in the upper right corner of the image.
{\bf b} (lower) contour map of the integrated intensity, overlaid on an
optical image extracted from the Digitized Sky Survey. Contour levels are
30, 60, \ldots, 180\,K\,km\,s$^{-1}$.
The region covered by the spectra shown in (a) is indicated by the dashed
polygon, and the cross shows the adopted centre and the direction
of the major and minor axis
\label{fig:n3628-1}
}
\end{figure*}
\setcounter{figure}{7}

The reference position for our observations was chosen on the IR intensity
peak (Rieke et al.\ \cite{rieke+80}; see Table \ref{tab:obs}). For the
data analysis (Figs.\ \arabic{figure}c and d) we shifted the
reference point by
$(\Delta {\rm R.A.}, \Delta{\rm Dec.}) = (-2''\!\!.0, 4''\!\!.0)$ onto the
measured CO(3--2) maximum, and the position angle was assumed to be
$70^{\circ}$, as in Tilanus et al.\ (\cite{tilanus+91}). This value lies
between the optical position angle ($65^{\circ}$)
and that determined by Neininger et al.\ (\cite{neininger+98})
from their interferometric $^{13}$CO map of the centre of M\,82.

The source size -- deconvolved by the
beam -- is measured to be $46'' \times 23''$, corresponding to a linear
scale of $710\,{\rm pc} \times 360\,{\rm pc}$ at the distance of M\,82.
This region is larger than in some other starburst galaxies in the
current sample, e.g.\ NGC\,253 or Maffei\,2. However, M\,82 is not a spiral
galaxy, but rather a small irregular, and rapid star formation could be
triggered over a large fraction of this object by its interaction with
M\,81.

\begin{figure*}
\resizebox{\hsize}{!}{\includegraphics*{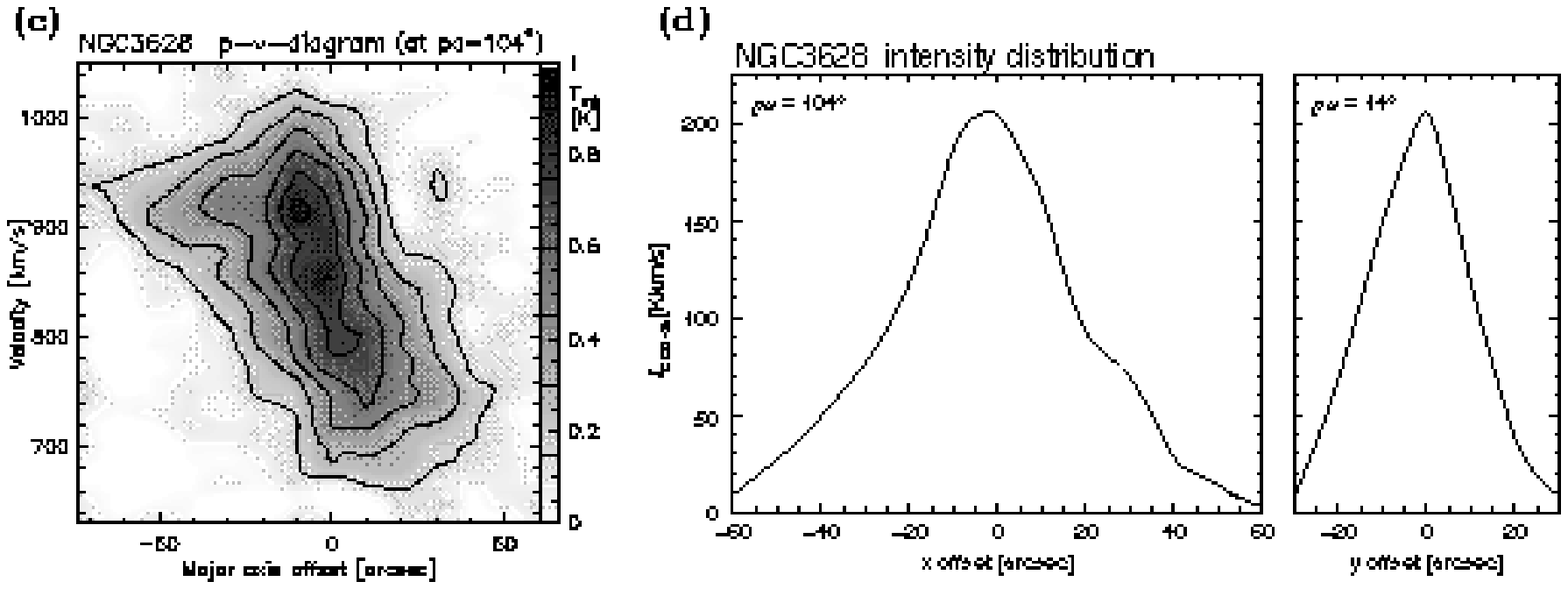}}
{\bf Fig.~8c and d.}~CO(3--2) data of NGC\,3628:
{\bf c} (left) position-velocity diagram along the major axis. Contours
are 0.15, 0.3, \ldots 0.9\,K.
{\bf d} (right) intensity distribution along the major and minor
axis, respectively
\label{fig:n3628-2}
\end{figure*}
\setcounter{figure}{8}

{\it Intensities and Line Ratios.}
M\,82 shows the strongest CO(3--2) emission of all objects in our sample.
Typical brightness temperatures are $T_{\rm mb} = 4 - 5\,{\rm K}$.
The total CO(3--2) line flux in the region covered by our map is
$F_{\rm CO(3-2)} = 14.63 \pm 0.40\,10^4\,{\rm Jy\,km\,s}^{-1}$;
the total power emitted in this line is
$P_{\rm CO(3-2)} = 16.4 \pm 0.5\,10^{30}\,{\rm W}$.
The line intensity measured in the central region is
$I_{\rm CO(3-2)} = 770 \pm 70\,{\rm K\,km\,s}^{-1}$. A direct comparison
with the values given by Wild et al.\ (\cite{wild+90}) and Tilanus et
al.\ (\cite{tilanus+91}) is difficult due to the different beam sizes.
The CO(1--0) data of Wild et al., however, were observed with the same
angular resolution as our CO(3--2) data and allow the determination
of line ratios. In the central region, we find a value of
$R_{3,1} = 1.0 \pm 0.2$, which is consistent with the results
presented by Tilanus et al.\ (\cite{tilanus+91}). At a distance of
$20''$ from the centre along the major axis, the line ratio drops
to $0.8 \pm 0.2$. Note the different reference position of Wild
et al.\ (\cite{wild+90}).
We will not discuss the physical properties of the gas in detail here,
but refer instead to the numerous multi-line investigations already
published (e.g.\ Wei{\ss} et al.\ \cite{weiss+01}).
We argue that a line ratio around unity -- in contrast to
other starburst galaxies -- may be due to optical depth effects
in the CO lines; see the publications
cited above for a dicussion including data from other isotopomers.

{\it Kinematics.}
A Gaussian fit to the central spectrum yields a line width
of $\Delta\,V \sim 230\,{\rm km\,s}^{-1}$. From the shape of the spectra,
however, it is clear that we see, even in the centre, two components
which represent a ring-like distribution. A two-component fit to the
central spectrum yields two separate lines with widths of about 130 and
100\,km\,s$^{-1}$ and at central velocities of
$V_0 \sim 180\,{\rm km\,s}^{-1}$ and $V_0 \sim 290\,{\rm km\,s}^{-1}$,
respectively. The kinematical structure of M\,82 as a whole is a bit more
complicated. The two components visible in the central spectrum extend to
radii of $\pm 30''$ on either side, where they rotate with a velocity of
$\pm 80\,{\rm km\,s}^{-1}$ relative to the systemic velocity of
225\,km\,s$^{-1}$. From the velocity field (not shown) we conclude further
that this rotation is in the plane of M\,82 (i.e.\ along the major axis).
The rotation above and below the plane seems to differ from this behaviour
significantly. Below the plane, the measured velocities of the CO(3--2)
emission are $< 200\,{\rm km\,s}^{-1}$, probably indicating an outflow of
highly excited molecular gas towards the observer in the southern halo
of M\,82.

\subsection{NGC\,3628}

The disturbed edge-on galaxy NGC\,3628 is a member of the well known
Leo triplet of galaxies and exhibits a starburst, which is probably
triggered by the interaction of NGC\,3628 with its neighbours NGC\,3623
and NGC\,3627.
It is viewed edge-on with an inclination of $89^{\circ}$, and its distance
is assumed here to 6.7\,Mpc (de Vaucouleurs \cite{devaucouleurs75})
which is consistent with previous studies of this object.
Despite its starburst properties, it was not observed in the higher CO
transitions before. Reuter et al.\ (\cite{reuter+91}) observed the
CO(2--1) line in this object, while Boiss\'e et al.\ (\cite{boisse+87})
observed several CO(1--0) spectra in the inner $2'$. Interferometric
CO observations were presented by Irwin \& Sofue (\cite{irwin+sofue96}).

{\it Morphology.}
The position of the CO maximum does not coincide with the optical centre,
a fact that was already noted by Boiss\'e et al.\ (\cite{boisse+87}) for
the CO(1--0) line.

Even if the CO(3--2) emission in NGC\,3628 shows a concentration towards
the central region, the intensity distribution along the major axis is
not as centrally peaked as for other starburst galaxies in our sample:
The deconvolved source size is $34'' \times 15''$, corresponding to a
linear size of $1.1\,{\rm kpc} \times 0.5\,{\rm kpc}$ at a distance
of 6.7\,Mpc. This slower intensity decrease along the major axis
indicates that violent star formation does not only occur in the very
centre, but rather in a central bar or disk which is somewhat
elongated along the major axis.

{\it Intensities and Line Ratios.}
The integrated CO(3--2) line flux measured from NGC\,3628 is
$F_{\rm CO(3-2)} = 2.68 \pm 0.14\,10^4\,{\rm Jy\,km\,s}^{-1}$. This
corresponds, at an assumed distance of 6.7\,Mpc, to a total power of
$P_{\rm CO(3-2)} = 1.31 \pm 0.07\,10^{31}\,{\rm W}$, which is of the
same order as for M\,82, but is emitted from a larger region. The maximum
intensity measured in the central region of NGC\,3628 is
$I_{\rm CO(3-2)} = 206 \pm 19\,{\rm K\,km\,s}^{-1}$.
From the CO(1--0) data of Boiss\'e et al.\ (\cite{boisse+87}), which
are observed with a similar beam size, we estimate a line ratio of
$R_{3,1} = 1.4 \pm 0.2$. This is, together with M\,83, the highest value
in the present sample.
From a CO(2--1) map from Reuter et
al.\ (\cite{reuter+91}), smoothed to $21''$ resolution, we can further
determine $R_{3,2} = 1.5 \pm 0.2$. For this galaxy the integrated intensity
of the CO(2--1) line is smaller than that of both the (1--0) and the
(3--2) transition in the central region. At radii larger than about
$30''$, both line ratios decrease to $1.0 \pm 0.2$, similar to the other
starburst objects in our sample.

\begin{figure*}
\resizebox{12cm}{!}{\includegraphics*{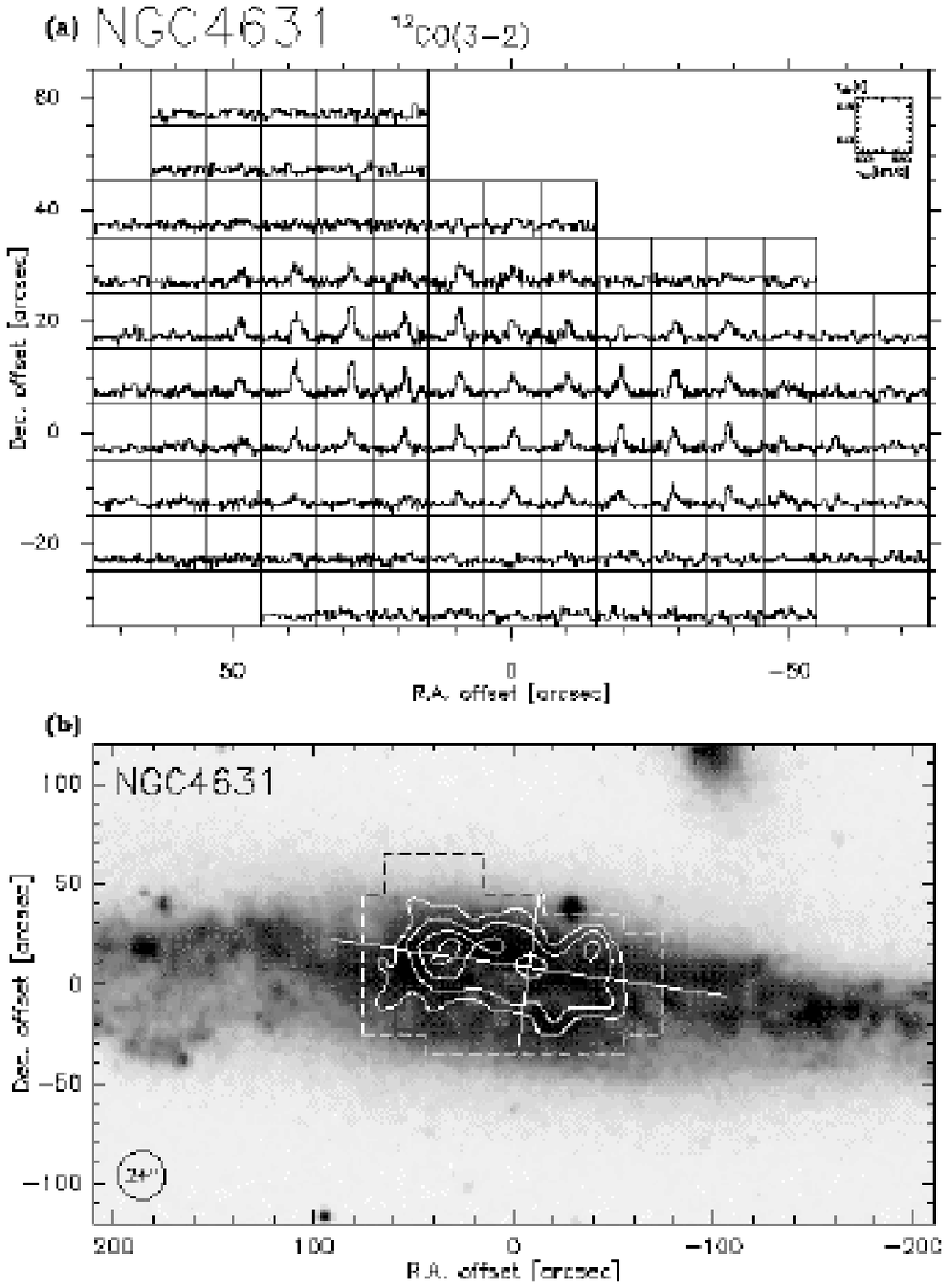}}
\hfill
\parbox[b]{55mm}{
{\bf Fig.~9a and b.}~CO(3--2) data of NGC\,4631:
{\bf a} (upper) raster map of the individual spectra, with (0,0)
corresponding to the position given in Table \ref{tab:obs}.
The scale of the spectra is indicated by the small box inserted
in the upper right corner of the image.
{\bf b} (lower) contour map of the integrated intensity, overlaid on an
optical image extracted from the Digitized Sky Survey. Contour levels are
11, 22, 33, 44\,K\,km\,s$^{-1}$.
The region covered by the spectra shown in (a) is indicated by the dashed
polygon, and the cross shows the reference position for data analysis
(see text) and the direction of the major and minor axis
\label{fig:n4631-1}
}
\end{figure*}
\setcounter{figure}{8}

In the centre, the high line ratio indicates a kinetic temperature of
the CO gas of $T > 60\,{\rm K}$, at molecular gas densities around
$n({\rm H}_2) \sim 10^{3.5}\,{\rm cm}^{-3}$.
The CO(3--2) spectrum presented by Mauersberger et
al.\ (\cite{mauersberger+99}) was taken at a position nearly $25''$ off
the centre toward the southeast. They estimated a line ratio of
$R_{3,1} \sim 0.2$, which is below our value in the disk, and is
not consistent with the starburst properties in the central region.
This again shows that an extended mapping of the target sources is
necessary for a detailed analysis of line ratios.

{\it Kinematics.}
The kinematics as traced by the CO(3--2) emission are dominated by a large
line width in the central region (see
Fig.\ \arabic{figure}c),
and by a large velocity gradient in the inner $20''$.
Both results suggest either fast rotation or non-circular velocities near
the centre. The latter may be a signature of a barred potential
in the inner part of NGC\,3628, feeding the starburst nucleus. The velocity
field (not shown) shows some velocity shear between the southern and
northern half. This may be a consequence of the interaction between
NGC\,3628 and its neighbours NGC\,3623 and NGC\,3627 (Rots \cite{rots78}).
 
\subsection{NGC\,4631}
\label{section:results_n4631}

\begin{figure*}
\resizebox{\hsize}{!}{\includegraphics*{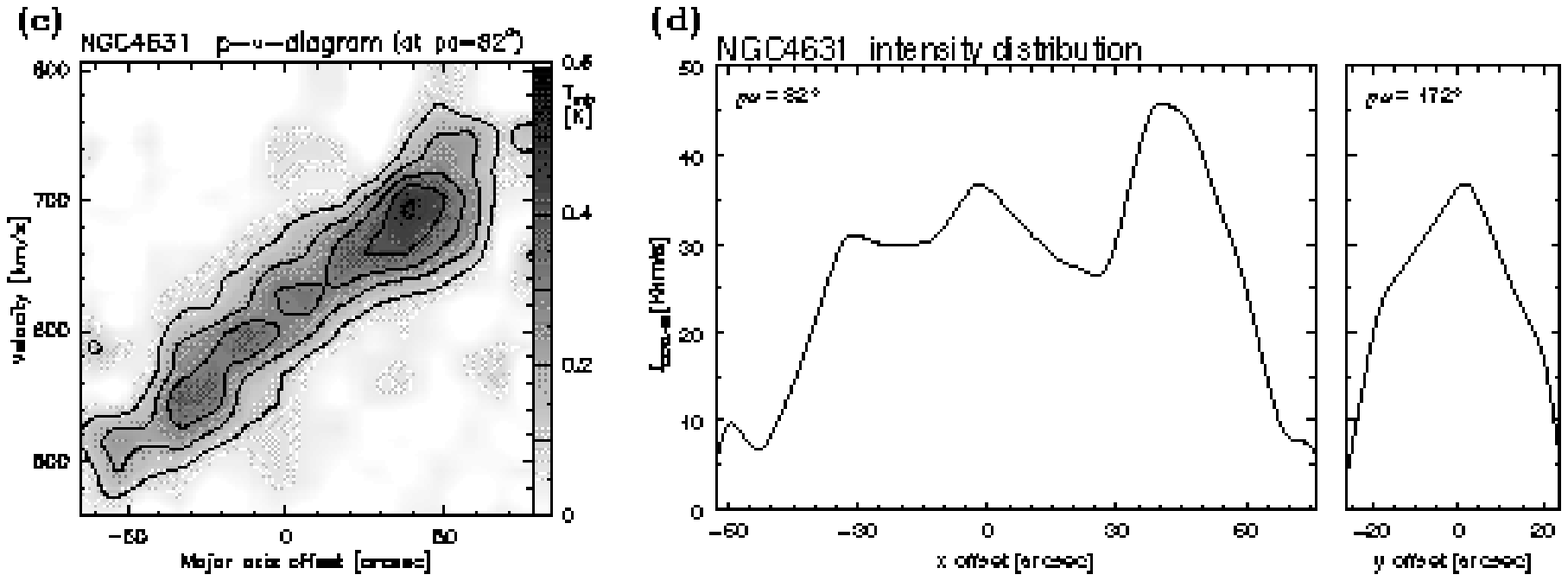}}
{\bf Fig.~9c and d.}~CO(3--2) data of NGC\,4631:
{\bf c} (left) position-velocity diagram along the major axis. Contours
are 0.1, 0.2, \ldots 0.5\,K.
{\bf d} (right) intensity distribution along the major and minor
axis, respectively.
Note that for figure parts {\bf c} and {\bf d} the zero position differs
from the position (0,0) in figure parts {\bf a} and {\bf b} by
$(\Delta {\rm R.A.}, \Delta{\rm Dec.}) = (-7''\!\!.6, 8''\!\!.0)$
\label{fig:n4631-2}
\end{figure*}
\setcounter{figure}{9}

NGC\,4631 is a disturbed, edge-on ($i \sim 86^{\circ}$) late-type galaxy.
It is a member of an interacting group of at least 3 galaxies which includes
also NGC\,4627 and NGC\,4656 (e.g.\ Combes \cite{combes78}). For a distance
of $D = 7.5\,{\rm Mpc}$ (consistent with earlier publications,
e.g.\ Golla \& Wielebinski \cite{golla+rw94}), the separation
of NGC\,4631 from its neighbours is about 60\,kpc and 4\,kpc, respectively.
As a consequence of this interaction, NGC\,4631
shows a disturbed appearance in the optical and is believed to exhibit a mild
starburst (Golla \& Wielebinski \cite{golla+rw94}). These authors have
observed the CO(1--0) and the CO(2--1) line transitions with the IRAM 30\,m
telescope. Recently Neininger \& Dumke (\cite{neininger+dumke99}) found
intergalactic cold dust (traced by continuum emission at 1.2\,mm)
in this galaxy group. This emission is connected with the \ion{H}{i}
spurs emerging from the disk (Rand \cite{rand94}), probably as a consequence
of the mentioned interaction. The CO(3--2) data we present here were
published as a smoothed raster map of spectra in
Wielebinski et al.\ (\cite{rw+md+cn99}).

{\it Morphology.}
The spatial distribution of the CO(3--2) emission
(Fig.\ \arabic{figure})
differs from the other starburst galaxies in our sample,
and shows that the starburst properties (if we follow the ``mild starburst''
classification of Golla \& Wielebinski \cite{golla+rw94}) are
different from starbursts occuring in the nuclei of galaxies.
Besides the central maximum with a velocity-integrated intensity of
$I_{\rm CO(3-2)} = 37 \pm 6\,{\rm K\,km\,s}^{-1}$ there are two further
maxima at radii of about $\pm 40''$, which indicate a disk- or
ring-like distribution of the CO gas emitting in the (3--2) line.
The eastern peak of this ring,
with an intensity of $I_{\rm CO(3-2)} = 46 \pm 7\,{\rm K\,km\,s}^{-1}$,
corresponds to the star-forming region CM\,67. Because of this
distribution, we cannot estimate a central peak size here. Thus we
estimate the size of the whole CO(3--2) emitting region. 
We find, assuming a disk-like morphology and after deconvolving in
$z$\/-direction, a size of $100'' \times 21''$. This corresponds to
$3.63\,{\rm kpc} \times 0.77\,{\rm kpc}$, of which, however, only the
latter value (the extent along the minor axis) can be compared with
the other galaxies in our sample. As noted by Wielebinski et
al.\ (\cite{rw+md+cn99}), the detected CO(3--2) emission is as extended
as the lower transitions observed by Golla \& Wielebinski
(\cite{golla+rw94}).

We note here that the intensity distributions
(Fig.\ \arabic{figure}d)
and the position-velocity diagram
(Fig.\ \arabic{figure}c)
are not relative to the position given in Table \ref{tab:obs}, but to
a reference position which is located at
$(\Delta {\rm R.A.}, \Delta{\rm Dec.}) = (-7''\!\!.6, 8''\!\!.0)$ in
Figs.\ \arabic{figure}a and b. This was done in order to center the
distributions on the CO(3--2) emission.
Interestingly, this reference position is even $\sim 5''$ northwest
of the central position
given by Golla \& Wielebinski (\cite{golla+rw94}), namely --
after conversion to J(2000) coordinates --
${\rm R.A.[2000]} = 12^{\rm h}42^{\rm m}07^{\rm s}\!\!.2$,
${\rm Dec.[2000]} = 32^{\circ}32'32''$.
Hence the CO(3--2) maxima seem to be slightly north of those for the
lower transitions. This is possibly connected to the outflow of interstellar
matter into the northern halo of NGC\,4631, either by varying optical
depth effects or changing gas properties with increasing distance from the
major axis.

{\it Intensities and Line Ratios.}
We measure a total line flux in NGC\,4631 of
$F_{\rm CO(3-2)} = 1.35 \pm 0.11\,10^4\,{\rm Jy\,km\,s}^{-1}$,
which corresponds to a total power of
$P_{\rm CO(3-2)} = 8.3 \pm 0.7\,10^{30}\,{\rm W}$ emitted
in the CO(3--2) line.
Due to the different locations of the emission maxima for the different
transitions, the line ratios vary for a constant galactocentric radius.
When comparing our data with the CO(1--0) data from
Golla \& Wielebinski (\cite{golla+rw94}), we find values of
$R_{3,1}$ between $0.9 \pm 0.2$ and $1.2 \pm 0.2$ in the central region and
between $0.6 \pm 0.2$ and $0.9 \pm 0.3$ close to the outer maxima.
The line ratio between the (2--1) and the (1--0) transition is about
$R_{2,1} = 0.8 - 0.9$ for all intensity peaks.
These line ratios indicate moderately warm molecular gas of
$\sim 30 - 50\,{\rm K}$. We conclude that NGC\,4631 is not an active
starburst galaxy. However, in the molecular ring, at radii of about
$\pm 40''$, there is some enhanced star formation taking place. This is
also suggested by the H$\alpha$ morphology (Golla et al.\ \cite{golla+96}).
In addition,
the huge radio halo visible at cm-wavelengths and the magnetic field
orientation (e.g.\ Hummel et al.\ \cite{hummel+91}; Golla \& Hummel
\cite{golla+hummel94}) point to some kind of activity in the
central region. This activity has probably been triggered by the
interaction of NGC\,4631 with its neighbours.

\begin{figure*}
\resizebox{\hsize}{!}{\includegraphics*{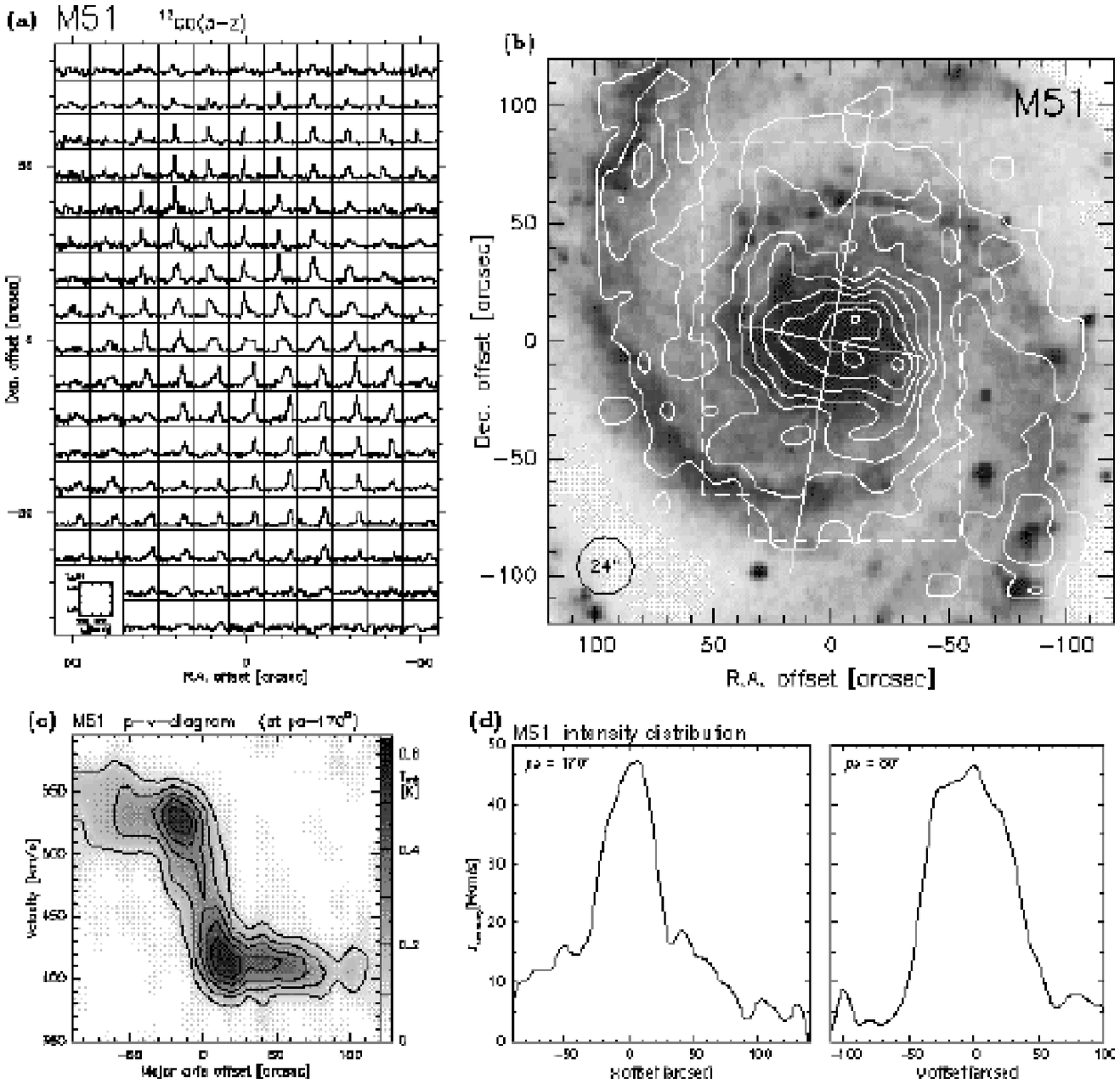}}
\caption{CO(3--2) data of M\,51:
{\bf a} (upper left) raster map of the individual spectra, with (0,0)
corresponding to the position given in Table \ref{tab:obs}.
The scale of the spectra is indicated by the small box inserted
in the lower left corner of the image. Only the inner part of the mapped
area is shown here.
{\bf b} (upper right) contour map of the integrated intensity, overlaid on an
optical image extracted from the Digitized Sky Survey. Contour levels are
6, 12, 18, \ldots, 48\,K\,km\,s$^{-1}$.
The region covered by the spectra shown in (a) is indicated by the dashed
polygon, and the cross shows the adopted centre and the direction
of the major and minor axis.
{\bf c} (lower left) position-velocity diagram along the major axis. Contours
are 0.1, 0.2, \ldots 0.6\,K.
{\bf d} (lower right) intensity distribution along the major and minor
axis, respectively.
Note that while only the spectra in the inner part of M\,51 are shown in (a),
the other figure parts include the whole data set as shown by
Wielebinski et al.\ (\cite{rw+md+cn99})
}
\label{fig:m51}
\end{figure*}

{\it Kinematics.}
The kinematics as obtained from the CO(3--2) emission do not show
significant differences from data from the lower transitions.
Rigid rotation can be seen out to radii of $40'' - 50''$. The line width
in the central region is narrow compared to the starburst galaxies in
our sample and also compared to the other edge-on objects, about
80\,km\,s$^{-1}$. This line width stays more or less constant along
the major axis of NGC\,4631. In the centre no fast rotating gas or
noncircular velocities can be seen. The velocity field (not shown) shows
further some kinematic anomalies in the northern half, also due to the
interaction of NGC\,4631 with its neighbours.

\subsection{M\,51}

M\,51 (NGC\,5194) is a grand-design face-on ($i = 20^{\circ}$)
spiral galaxy at a distance of $D = 9.6\,{\rm Mpc}$ (Sandage \& Tammann
\cite{sandage+tammann75}).
It has already been subject to several studies of the molecular line
emission (e.g.\ Nakai et al.\ \cite{nakai+94}; Garc\'{\i}a-Burillo et
al.\ \cite{garcia+93}, and references therein). The data for this
galaxy were shown by Wielebinski et al.\ (\cite{rw+md+cn99}) to
demonstrate that the CO(3--2) emission is also found in the disks
of galaxies. Thus the physical conditions in the spiral arms also
lead to a significant excitation of the $J=3$ level of CO. In the
present paper we only show the spectra of the inner disk of M\,51.
The distribution of the CO(3--2) emission in the spiral arms can be
seen in Fig.\ \ref{fig:m51}b, which shows an overlay of our
CO(3--2) intensity map, as obtained from the complete data set,
on an optical image.

\begin{figure*}
\resizebox{\hsize}{!}{\includegraphics*{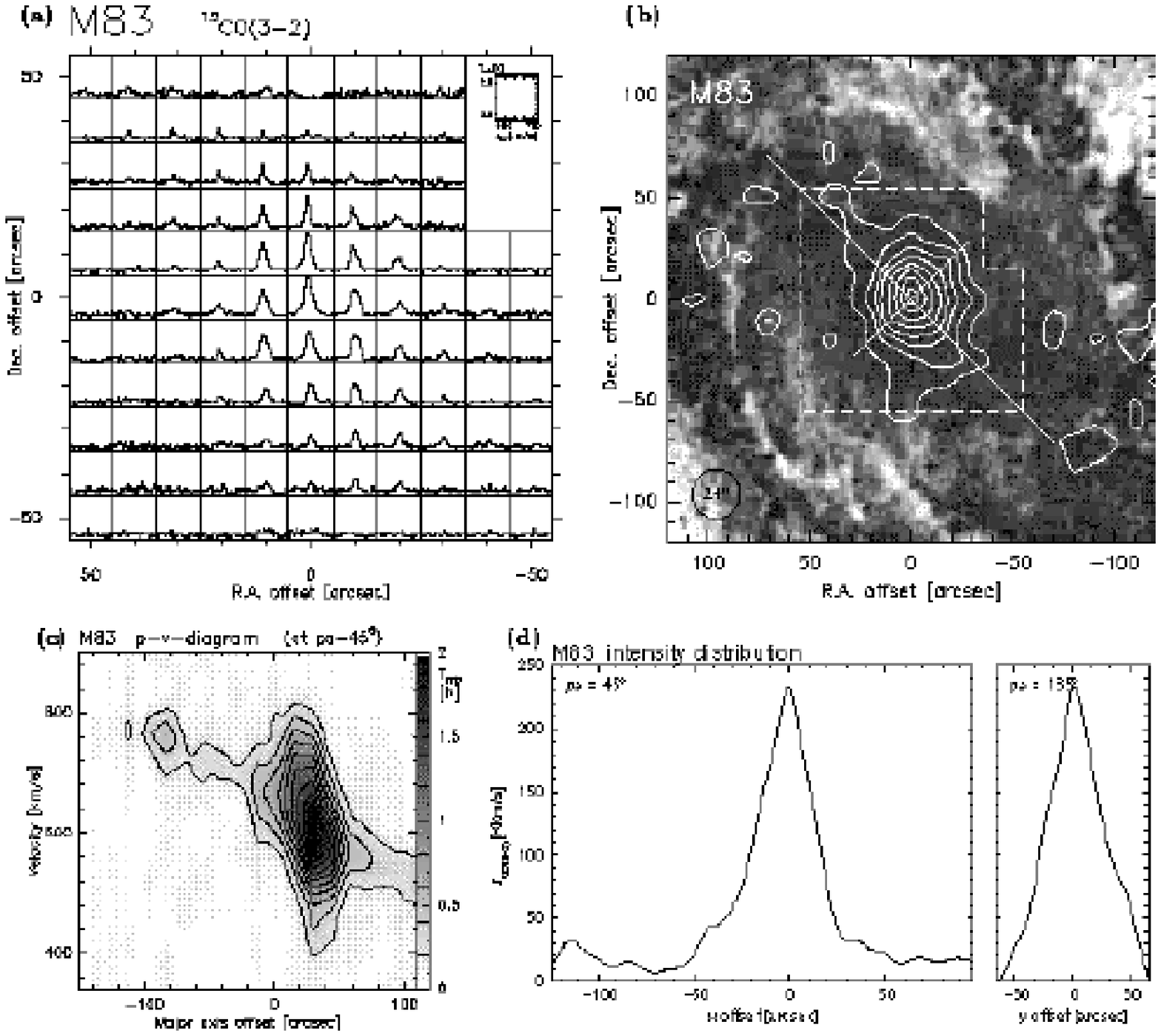}}
\caption{CO(3--2) data of M\,83:
{\bf a} (upper left) raster map of the individual spectra, with (0,0)
corresponding to the position given in Table \ref{tab:obs}.
The scale of the spectra is indicated by the small box inserted
in the upper right corner of the image. Only the inner part of the mapped
area is shown here.
{\bf b} (upper right) contour map of the integrated intensity, overlaid on an
optical image extracted from the Digitized Sky Survey. Contour levels are
25, 50, 75, 100, 130, \ldots, 220\,K\,km\,s$^{-1}$.
The region covered by the spectra shown in (a) is indicated by the dashed
polygon, and the cross shows the adopted centre and the direction
of the major and minor axis.
{\bf c} (lower left) position-velocity diagram along the major axis. Contours
are 0.2, 0.4, \ldots 1.8\,K.
{\bf d} (lower right) intensity distribution along the major and minor
axis, respectively.
Note that while only the spectra in the inner part of M\,83 are shown in (a),
the other figure parts include the whole data set (see text)
}
\label{fig:m83}
\end{figure*}

\begin{figure*}
\resizebox{\hsize}{!}{\includegraphics*{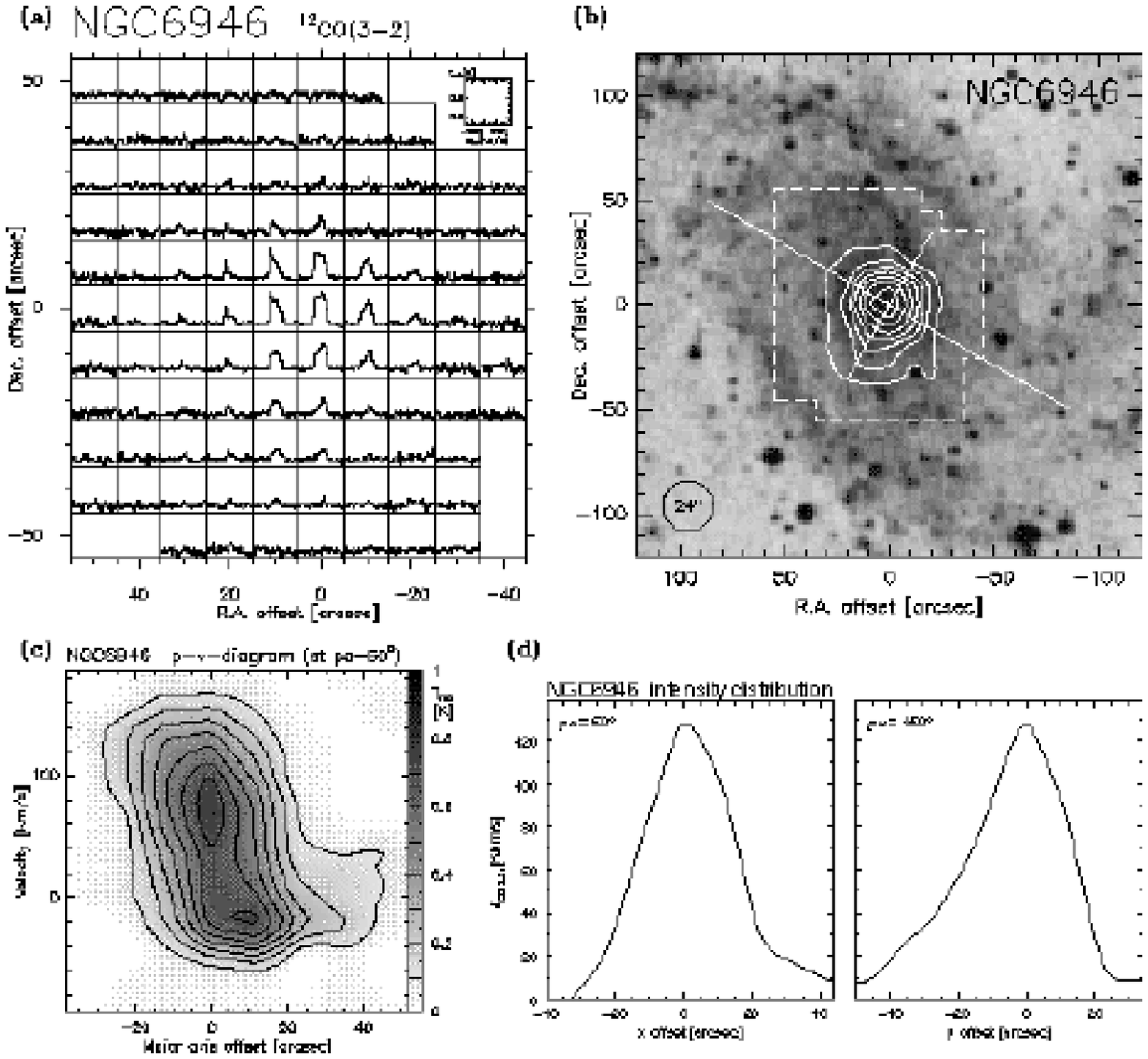}}
\caption{CO(3--2) data of NGC\,6946:
{\bf a} (upper left) raster map of the individual spectra, with (0,0)
corresponding to the position given in Table \ref{tab:obs}.
The scale of the spectra is indicated by the small box inserted
in the upper right corner of the image.
{\bf b} (upper right) contour map of the integrated intensity, overlaid on an
optical image extracted from the Digitized Sky Survey. Contour levels are
16, 32, 48, \ldots, 128\,K\,km\,s$^{-1}$.
The region covered by the spectra shown in (a) is indicated by the dashed
polygon, and the cross shows the adopted centre and the direction
of the major and minor axis.
{\bf c} (lower left) position-velocity diagram along the major axis. Contours
are 0.1, 0.2, \ldots 0.8\,K.
{\bf d} (lower right) intensity distribution along the major and minor
axis, respectively
}
\label{fig:n6946}
\end{figure*}

{\it Morphology.}
Even though the CO(3--2) emission is seen in the spiral arms, emission
from the centre dominates, with a maximum intensity of
$I_{\rm CO(3-2)} = 47 \pm 7\,{\rm K\,km\,s}^{-1}$. This is a factor 3
brighter than the maxima in the spiral arms. Interestingly the extent
of the central peak is larger along the minor axis than along the major
axis (the position angle can easily be estimated from the isovelocity
contours in the velocity field); the deconvolved source size is
$47'' \times 76''$. This corresponds to
$2.2\,{\rm kpc} \times 3.5\,{\rm kpc}$ at a distance of 9.6\,Mpc.
Another difference between M\,51 and most other objects is the smooth
transition between the line emitting central region and the spiral arms.
For most galaxies in this survey the CO(3--2) emission drops to
almost zero outside the central area, before it rises again within
the spiral arms, as in M\,83 (Fig.\ \ref{fig:m83}b) or NGC\,6946
(Fig.\ \ref{fig:n6946}b and Nieten et al.\ \cite{nieten+99}).
This connection between the highly excited CO gas in the centre and the
disk is further seen only in NGC\,891.

{\it Intensities and Line Ratios.}
The total line flux measured for the central peak of M\,51 is
$F_{\rm CO(3-2)} = 1.6 \pm 0.2\,10^4\,{\rm Jy\,km\,s}^{-1}$,
the total power radiated from the central region in the CO(3--2) line
$P_{\rm CO(3-2)} = 1.6 \pm 0.2\,10^{31}\,{\rm W}$. The latter value is at the
higher end of the range represented by our galaxy sample. When we normalize
this value by the size of the emitting region, M\,51 shows a rather
average result.

Wielebinski et al.\ (\cite{rw+md+cn99}) compared the present data with
the CO(1--0) observations of Nakai et al.\ (\cite{nakai+94}) and
determined realtively low line ratios of $R_{3,1} = 0.5 \pm 0.2$ in the
central region and $R_{3,1} = 0.7 \pm 0.2$ in the spiral arms. They
observed, however, even higher values in the outermost spiral arms
($R_{3,1} = 1.2 \pm 0.35$). The ratio between the (2--1) and the (1--0)
transition was estimated by Garc\'{\i}a-Burillo et al.\ (\cite{garcia+93})
to be almost constant over their map, $R_{2,1} \sim 0.8$ (with a similar
angular resolution), but slightly higher in the centre. This leads to
line ratios of $R_{3,2} \sim 0.65$ in the centre and $ \sim 0.9$
in the spiral arms. This result is not consistent with that determined by
Garc\'{\i}a-Burillo et al.\ (\cite{garcia+93}) from a few positions
observed in the CO(3--2) line with the 30\,m telescope.
The difference, however, may be
explained by the smaller beam size and the error beam of the 30\,m telescope.
The estimated values for $R_{3,1}$ are -- beside NGC\,891 -- the lowest in
our sample, and the inferred properties of the molecular gas are thus
similar to NGC\,891. Densities are $\sim 10^3\,{\rm cm}^{-2}$ or below;
and temperatures are in the range $20 - 40\,{\rm K}$

{\it Kinematics.}
The kinematical structure of the inner disk of M\,51, as revealed by the
position-velocity diagram and the velocity field, is similar
to the CO(2--1) data from 
Garc\'{\i}a-Burillo et al.\ (\cite{garcia+93}). Along the major axis the gas
rotates rigidly out to a radius of $15'' - 20''$, and with a constant
velocity further out. In the $p$\/-$v$\/-diagram (Fig.\ \ref{fig:m51}c),
the second and third temperature maximum, which are
clearly seen in the CO(2--1) data at $r \sim 50''$ and $r \sim 120''$
respectively, are less pronounced (or even missing) in the CO(3--2) emission.

From a Gaussian fit to the central profile, we obtain a line
width of $\Delta\,V \sim 120\,{\rm km\,s}^{-1}$. Since the profile is
composed of two components, an appropriate fit yields line
widths of $\sim 60\,{\rm km\,s}^{-1}$ for these components
at central velocities of 430 and 500\,km\,s$^{-1}$.

\subsection{M\,83}

M\,83 is a barred galaxy at a distance of $D = 4.7\,{\rm Mpc}$
(Tully \cite{tully88}). With an inclination of $i = 24^{\circ}$
(Comte \cite{comte81}) it is oriented relatively face-on. It is generally
believed to undergo a nuclear starburst which is triggered by gas infall
along the bar potential (Talbot et al.\ \cite{talbot+79}).

Observations of the CO(1--0) line were performed basically in the spiral
arms of this large nearby galaxy. Handa et al.\ (\cite{handa+90}) made
single-dish observations of the central region with an angular resolution
of $16''$ HPBW\@. Petitpas \& Wilson (\cite{petitpas+wilson98}) observed
the inner $\pm 20''$ in the (3--2) and the (4--3) transition of CO.

{\it Morphology.}
M\,83 was also observed at larger radii from the centre within the spiral
arms during our observing sessions (Thuma et al., in prep.), but we will
restrict the discussion here to the emission from the central part of this
galaxy. The line
temperature reaches a maximum of $T_{\rm mb} \sim 2.3\,{\rm K}$, a value
which is consistent with the JCMT observations of the CO(3--2) line
of Petitpas \& Wilson (\cite{petitpas+wilson98}), considering their smaller
telescope beam (and higher noise). The emission is concentrated to the
nucleus, but also slightly elongated along the bar.
The deconvolved extent of the central peak is $25'' \times 23''$,
corresponding to a linear size of $580\,{\rm pc} \times 520\,{\rm pc}$
at a distance of 4.7\,Mpc.

{\it Intensities and Line Ratios.}
The maximum intensity measured on the central peak is
$I_{\rm CO(3-2)} = 234 \pm 14\,{\rm K\,km\,s}^{-1}$. This is also consistent
with the data of Petitpas \& Wilson (\cite{petitpas+wilson98}) -- note that
we are not able to resolve the double peak structure along the bar given
the spatial resolution of our data. Even if we account for the slightly
different reference position used by
Mauersberger et al.\ (\cite{mauersberger+99})
and the large intensity gradient near the centre, the value of
$I_{\rm CO(3-2)} = 126\,{\rm K\,km\,s}^{-1}$ given by these authors
underestimates the emission by about 25\,\%.
The total line flux from the region shown in Fig.\ \ref{fig:m83}a can
be determined to
$F_{\rm CO(3-2)} = 3.12 \pm 0.15\,10^4\,{\rm Jy\,km\,s}^{-1}$, which
corresponds to a total power of
$P_{\rm CO(3-2)} = 7.5 \pm 0.4\,10^{30}\,{\rm W}$.
From Fig.\ \ref{fig:m83}d we estimate that a fraction of 10 -- 20\,\% of
this emission does not originate in the central peak, but in an underlying
molecular bar. 

Handa et al.\ (\cite{handa+90}) measured a CO(1--0) intensity near the
centre of $I_{\rm CO(1-0)} \sim 180\,{\rm K\,km\,s}^{-1}$ (converted to
$T_{\rm mb}$). Given the smaller beam size of their data, this can be used
to estimate a lower limit for the line ratio of $R_{3,1} > 1.2$. However,
since these authors give a fully sampled CO(1-0) map, we can simulate
a measurement with a larger beamsize by averaging the intensity values
in the inner $30''$ of their map with an appropriate weighting. In this
way we get a line ratio of $R_{3,1} = 1.4 \pm 0.3$ (the large error accounts
for the uncertainties of this method). In the disk, at radii between $30''$
and $50''$, the line ratio goes down to $0.9 \pm 0.3$.
Therefore M\,83 seems to have similar gas properties to most other
galaxies in our sample.

{\it Kinematics.}
The $p$\/-$v$\/-diagram along the major axis (Fig.\ \ref{fig:m83}c) shows a
superposition of two kinematical structures. In the inner $\pm 15''$ we see
a large velocity gradient from the systemic velocity of 510\,km\,s$^{-1}$
up to a rotational velocity of $\sim \pm 80\,{\rm km\,s}^{-1}$ relative to
$v_{\rm sys}$. This signature in the $p$\/-$v$\/-diagram may be due to the
$x_2$ orbits caused by the potential of the stellar bar which can be seen
in optical images. A second kinematical system, probably due to the $x_1$
orbits of the mentioned bar, shows a smaller velocity
gradient along the line defined by the position angle of M\,83, reaching
the same rotational velocity, but at a distance of about $\pm 120''$ from
the centre. Near the centre we measure a line width of
$\Delta\,V \sim 100\,{\rm km\,s}^{-1}$.

\subsection{NGC\,6946}

NGC\,6946 is a nearby grand-design spiral galaxy, oriented nearly face-on
($i = 30^{\circ}$) and classified as SAB(rs)cd. Distance estimates for
NGC\,6946 range from 3.2 to 11\,Mpc. De Vaucouleurs (\cite{devaucouleurs79})
gives 5.1\,Mpc, while Ferguson et al.\ (\cite{ferguson+98}) use 5.3\,Mpc.
We will assume a distance of 5.2\,Mpc here.

NGC\,6946 is believed to undergo a
moderate starburst in its nucleus (Turner \& Ho \cite {turner+ho83}).
It has already been studied at several wavelengths, but due to its
large optical size there is no complete map in one of the CO transitions.
While Sofue et al.\ (\cite{sofue+88}) and Weliachew et
al.\ (\cite{weliachew+88}) observed the CO(1--0) transition only in the
central region, Young \& Scoville (\cite{young+scoville82}) obtained some
data along two orthogonal strips. Casoli et al.\ (\cite{casoli+90}) added
some information about two regions in spiral arms $\pm 150''$ from the
centre in the CO(1--0) and (2--1) lines. The first results for higher CO
transitions were presented by Nieten et al.\ (\cite{nieten+99}). These
authors observed the CO(4--3) line in the inner part, using the HHT. While
a few selected CO(3--2) spectra were already shown in that publication,
here we present the full map of the inner disk of NGC\,6946. We note that
in this galaxy CO(3--2) emission is also found in the spiral arms, more
than $3'$ away from the centre. More extensive mapping of these off-centre
regions in the CO lines has been performed by Walsh et al.\ (in prep.).

{\it Morphology.}
Emission from the inner disk of NGC\,6946 is concentrated to the nucleus.
This result was also found by Weliachew et al.\ (\cite{weliachew+88}),
with a similar beam size, in the CO(1--0) line. The CO(3--2) data show a
slightly larger extent to the south-southeast and a large intensity
gradient in the opposite direction (see Figs.\ \ref{fig:n6946}\,b and d).
However, the distribution is not point-like -- the deconvolved source
extent is $19'' \times 21''$ along the major and minor axis respectively,
corresponding to $480\,{\rm pc} \times 530\,{\rm pc}$ at the distance of
NGC\,6946. 

The brightness temperatures measured in the central region are about
$T_{\rm mb} = 0.8\,{\rm K}$. The CO(3--2) spectrum shown by Mauersberger
et al.\ (\cite{mauersberger+99}) is a factor of 2 weaker. Their observing
position, however, is located at $(-5'', +10'')$ relative to our central
coordinates, hence the results are not inconsistent.

\begin{table*}
\caption[]{General source properties obtained from the literature.
Basic references are de Vaucouleurs et al.\ (\cite{rc3}) and Young et
al.\ (\cite{young+89}); others are cited in the text.
}
\label{tab:basic}
\begin{tabular}{llccccccccc}
\hline
Source  & Type & $D$ & beam size & $pa$ & Incl.
	& $D_{25}$ & IR color & $L_{\rm IR}$ & $L_{\rm IR}/L_{\rm B}$
	& $ M_{\rm \ion{H}{i}} $ \\
 & & [Mpc] & $24'' \widehat{=}$ [pc] & [$^{\circ}$] & [$^{\circ}$]
	& [$'$] & $S_{60}/S_{100}$ & $[\times 10^9 L_{\odot}]$ &
	& $ [\times 10^9 M_{\odot}] $ \\
\hline
NGC\,253  & SAB(s)c & 2.5 & 290 & $52^{\circ}$ & $78^{\circ}$
	& 27.5 & 0.66 & 15.10 & 1.42 & 2.57 \\
NGC\,278  & SAB(rs)b & 12.4 & 1440 & $55^{\circ}$ & ?
	& 2.1 & 0.63 &  8.73 & 0.88 & 1.16  \\
NGC\,891  & SA(s)b & 9.6 & 1120 & $23^{\circ}$ & $88^{\circ}$
	& 13.5& 0.41 & 19.32 & 1.16 & 3.73  \\
Maffei 2  & SAB(rs)bc & 5.0 & 580 & $26^{\circ}$ & $66^{\circ}$
	& 1? & --  &   -- & --  &  -- \\
IC\,342   & SAB(rs)cd & 4.5 & 520 & $25^{\circ}$ & $25^{\circ}$
	& 21.4& 0.46 &  2.26 & 0.62 & 18.2 \\
NGC\,2146 & SB(s)ab pec & 17.0 & 1980 & $135^{\circ}$ & $65^{\circ}$
	& 6.0 & 0.83 & 85.74 & 3.10 & 8.49 \\
M\,82     & I0 & 3.2 & 370 & $70^{\circ}$ & $81^{\circ}$
	& 11.2 & 1.02 & 29.74 & 5.74 & 1.30 \\
NGC\,3628 & SB pec & 6.7 & 780 & $104^{\circ}$ & $89^{\circ}$
	& 14.8 & 0.48 &  6.31 & 0.55 & 2.81 \\
NGC\,4631 & SB(s)d sp & 7.5 & 870 & $82^{\circ}$ & $86^{\circ}$
	& 15.5 & 0.53 & 11.90 & 0.56 &10.1 \\
M\,51     & SA(s)bc pec & 9.6 & 1120 & $170^{\circ}$ & $20^{\circ}$
	& 11.2 & 0.46 & 30.90 & 0.60 & 4.71 \\
M\,83     & SAB(s)c & 4.7 & 550 & $45^{\circ}$ & $24^{\circ}$
	& 12.9 & 0.54 & 14.98 & 0.65 & 8.10 \\
NGC\,6946 & SAB(rs)cd & 5.2 & 610 & $60^{\circ}$ & $30^{\circ}$
	& 11.5 & 0.51 & 10.80 & 0.64 & 5.32 \\
\hline
\end{tabular}
\end{table*}

{\it Intensities and Line Ratios.}
The total CO(3--2) line flux from the central region of NGC\,6946 is
$F_{\rm CO(3-2)} = 1.42 \pm 0.07\,10^4\,{\rm Jy\,km\,s}^{-1}$,
the total power emitted in this line is
$P_{\rm CO(3-2)} = 4.2 \pm 0.2\,10^{30}\,{\rm W}$.
From our data we calculate a maximum integrated line intensity of
$I_{\rm CO(3-2)} = 129 \pm 10\,{\rm K\,km\,s}^{-1}$. By comparing this
value with the CO(1-0) data from Weliachew et al.\ (\cite{weliachew+88}),
we estimate a line ratio of $R_{3,1} = 1.3 \pm 0.2$ in the central region.
Further we find $R_{3,1} = 1.0 \pm 0.2$ at a distance of $30''$ from
the centre.
These values are again similar to those for other starburst galaxies
in our sample.

{\it Kinematics.}
The kinematics in the inner region of NGC\,6946, as traced by CO(3--2),
show no differences from the properties found by Weliachew et
al.\ (\cite{weliachew+88}) in the CO(1--0) line. We see a steep
velocity gradient and a large line width in the centre. From a Gaussian fit
we obtain a line width of $\Delta\,V \sim 150\,{\rm km\,s}^{-1}$ for the
central spectrum. 

\section{Discussion and conclusions}
\label{section:discussion}

\subsection{General properties and activity}

For the target galaxies we collect some parameters from
the literature that are of interest when comparing the galaxies in
Table \ref{tab:basic}. These include some results from IR measurements
which can be used as a tracer for the star forming activity. However,
care must be taken because these values were obtained using IRAS and are
therefore {\it global} values, while the CO(3--2) data usually refer to
the central regions of the target galaxies. These data include the IR color,
$S_{60}/S_{100}$, which is the ratio of the measured flux densities at
60 and 100\,$\mu$m respectively. This is correlated with the dust temperature
in the galaxies. The IR luminosity, $L_{\rm IR}$, also depends on the
dust temperature, but is, according to Young et al.\ (\cite{young+89}),
more closely related to the total molecular mass as given by CO(1--0)
measurements. This explains the high IR luminosity for spiral galaxies
which form stars because of their large amount of molecular gas
(e.g.\ NGC\,891 and M\,51) and not because of peculiar physical properties.
We also give the assumed distance to each galaxy in
Table \ref{tab:basic}, since some determined parameters (luminosities,
masses, sizes) are distance dependent.

\subsection{Kinematics}

As already mentioned, the CO(3--2) line traces a different component of the
molecular gas in galaxies than the lower transitions. The kinematic
properties of this component may differ from the bulk of cold molecular gas,
and from other ISM components.

In general, the largest line widths are seen in the edge-on galaxies. Here
they reflect the rigid rotation of the inner gas disk, as well as possible
non-circular motions. In our sample, the objects classified as starbursts
show the largest linewidths: up to 280\,km\,s$^{-1}$ for NGC\,2146. The
normal galaxy NGC\,891 and NGC\,4631 (sometimes called a ``mild starburst'')
show only linewidths of 100\,km\,s$^{-1}$ and 80\,km\,s$^{-1}$ in their
central regions, respectively.
Note that these values are not corrected for inclination, otherwise the
difference between NGC\,2146 and the other galaxies would be even larger.

For galaxies which are oriented close to face-on, the measured linewidth is
caused by the velocity dispersion of the gas. This dispersion is
expected to depend on the star-forming activity. The smallest values are
measured for NGC\,278 (40\,km\,s$^{-1}$) and IC\,342 (58\,km\,s$^{-1}$),
while all other face-on objects show values between 100 and 150\,km\,s$^{-1}$.

\subsection{Extended emission}

One of the most important results of this investigation is that the CO(3--2)
emission -- and hence the warm and dense gas -- is not restricted to the
nuclei of the galaxies as suggested earlier
(e.g.\ Mauersberger et al.\ \cite{mauersberger+96}), a fact
already emphasized by Wielebinski et al.\ (\cite{rw+md+cn99}). This shows the
necessity of extended mapping also in the higher CO line transitions.
To better quantify the extent of the emission and to set the CO(3--2)
morphology in relation to the general properties and types of the
observed galaxies, we investigate the extended gas in more detail.

From the intensity distribution we estimated the source size by fitting
a Gaussian to the central emission peak, and deconvolving it by the
(very simple) assumption that the intrinsic (unconvolved) intensity
distribution is also Gaussian. For most of the galaxies the central
emission peak could be reasonably well fitted by a Gaussian
with an error for the estimated width of about
$\pm 2''$. For NGC\,278 and NGC\,4631, the extent was estimated by the
assumption that the emission is distributed disk-like. Here the error
is about $\pm 4''$.
See also the individual subsections of
Sect.\ \ref{section:results} for details. The results are listed in
Table \ref{tab:results}, for the measured (in $''$) and the deconvolved
(in kpc) size.

The sizes of the central emission peak vary between 300\,pc and more than
3\,kpc between the galaxies. There is no correlation between the peak
size and other properties of the galaxies. Among the edge-on galaxies,
however, the disturbed starburst galaxy NGC\,2146 shows a very large
extent along the minor axis ($\sim 1.4\,{\rm kpc}$), which is probably
connected to the peculiar optical appearance and the high star formation
in this galaxy.

\subsection{Total fluxes}

\begin{table*}
\caption[]{Source properties as obtained from the CO(3--2) data.
In cols.\ 2 to 6 values are given for the central emission peak.
The source extent along the major and minor axis is given as measured size
(in $''$, col.\ 4) and already deconvolved by a Gaussian beam with
$HPBW = 24''$ (in kpc, col.\ 5). See text for further details.}
\label{tab:results}
\tabcolsep4pt
\begin{tabular}{lccccccc}
\hline
Source  & $F_{\rm CO(3-2)}$ & $P_{\rm CO(3-2)}$ &
	$\Theta_{\rm x} \times \Theta_{\rm y}$ & $l_{\rm x} \times l_{\rm y}$
	& $\Sigma_{\rm CO(3-2)}$
	& line ratios & $\Delta\,V_{\rm centre}$ \\
 & [$\times 10^4$\,Jy\,km\,s$^{-1}$] & [$\times 10^{30}\,{\rm W}$] 
	& [$'' \times ''$] & [kpc $\times$ kpc]
	& [$\times 10^{30}$\,W\,kpc$^{-2}$]
 	& $R_{3,1}$ (centre/disk) & [km\,s$^{-1}$] \\
\hline
NGC\,253  &  $8.74 \pm 0.30$ &  $6.0 \pm 0.2$ & $43'' \times 28''$
	& $0.43 \times 0.18 \pm 0.03$ & $32.4 \pm 3.7$
        & $( 0.8 / 0.5 ) \pm 0.1$ & 200 \\
NGC\,278  &  $0.37 \pm 0.03$ &  $6.2 \pm 0.5$ & $45'' \times 46''$
	& $2.71 \times 2.77 \pm 0.24$ & $0.83 \pm 0.14$
        & $0.8 \pm 0.2$ &  40 \\
NGC\,891  &  $0.47 \pm 0.06$ &  $4.7 \pm 0.6$ & $30'' \times 25''$
	& $0.80 \times 0.36 \pm 0.10$ & $7.3 \pm 2.0$
        & $( 0.4 / 0.5 ) \pm 0.1$ & 100 \\
Maffei 2  &  $1.87 \pm 0.08$ &  $5.1 \pm 0.3$ & $40'' \times 32''$
	& $0.78 \times 0.52 \pm 0.05$ & $8.4 \pm 1.5$
        & $( 1.3 / 0.8 ) \pm 0.2$ & 150 \\
IC\,342   &  $2.19 \pm 0.06$ &  $4.9 \pm 0.2$ & $45'' \times 34''$
	& $0.84 \times 0.52 \pm 0.05$ & $11.2 \pm 1.5$
        & $( 1.3 / 1.0 ) \pm 0.1$ &  60 \\
NGC\,2146 &  $1.94 \pm 0.14$ & $61.4 \pm 4.5$ & $37'' \times 29''$
	& $2.36 \times 1.36 \pm 0.17$ & $11.0 \pm 1.6$
        & $( 1.3 / 1.0 ) \pm 0.2$ & 280 \\
M\,82     & $14.63 \pm 0.40$ & $16.4 \pm 0.5$ & $52'' \times 33''$
	& $0.71 \times 0.36 \pm 0.04$ & $32.5 \pm 3.1$
        & $( 1.0 / 0.8 ) \pm 0.2$ & 230 \\
NGC\,3628 &  $2.68 \pm 0.14$ & $13.1 \pm 0.7$ & $41'' \times 29''$
	& $1.09 \times 0.50 \pm 0.07$ & $11.0 \pm 1.3$
        & $( 1.4 / 1.0 ) \pm 0.2$ & 220 \\
NGC\,4631 &  $1.35 \pm 0.11$ &  $8.3 \pm 0.7$ & $100'' \times 32''$
	& $3.63 \times 0.77 \pm 0.15$ & $0.63 \pm 0.07$
        & $( 1.0 / 0.7 ) \pm 0.2$ &  80 \\
M\,51     &  $1.60 \pm 0.20$ & $16.1 \pm 2.0$ & $53'' \times 80''$
	& $2.21 \times 3.55 \pm 0.10$ & $2.05 \pm 0.25$
        & $( 0.5 / 0.7 ) \pm 0.2$ & 130 \\
M\,83     &  $3.12 \pm 0.15$ &  $7.5 \pm 0.4$ & $35'' \times 33''$
	& $0.58 \times 0.52 \pm 0.05$ & $24.9 \pm 3.9$
        & $( 1.4 / 0.9 ) \pm 0.3$ & 100 \\
NGC\,6946 &  $1.42 \pm 0.07$ &  $4.2 \pm 0.2$ & $31'' \times 32''$
	& $0.48 \times 0.53 \pm 0.06$ & $16.5 \pm 3.3$
        & $( 1.3 / 1.0 ) \pm 0.2$ & 150 \\
\hline
\end{tabular}
\tabcolsep6pt
\end{table*}

Since we have obtained extended maps of the observed galaxies, it is
possible to calculate the total flux integrated over the mapped region.
Since this region contains in any case the central emission peak, some
conclusions about the total energy radiated in the CO(3--2) line are
possible.

The total flux is proportional to the velocity-integrated intensity and
the mapped area on the sky,
\begin{equation}
F_{\rm CO(3-2)} = {2\,k\,\nu^2 \over c^2}\,\int I_{\rm CO(3-2)}\,d\Omega\;,
\end{equation}
and is a measure of the total amount of highly excited CO gas, weighted
with the distance. The effect of the latter can be eliminated by comparing
the power radiated in the CO(3--2) line. From the values given in
Table \ref{tab:results}, we find some indication that the spiral galaxies
whose Hubble type classification includes ``peculiar'' (NGC\,2146,
NGC\,3628, and M\,51; see col.\ 2 in Table \ref{tab:basic}), as well
as M\,82, show an enhanced output in the CO(3--2) line. This means that
disturbances, e.g.\ due to encounters with neighbouring galaxies, lead
to larger amounts of highly excited molecular gas. Interestingly, this
conclusion is not valid for starburst galaxies in general:
Maffei\,2 and NGC\,253 show a similar value of $P_{\rm CO(3-2)}$ to
the normal galaxy NGC\,891 or NGC\,4631, a ``classical'' interacting galaxy.
Note that it is difficult to compare $P_{\rm CO(3-2)}$ for NGC\,278
with the other objects, since in NGC\,278 our map covers the entire
galaxy.

We can obtain a different picture when we normalize the power emitted in the
CO(3--2) line to the area from which this power is emitted. Thus we
define the frequency-integrated radiation surface density by
\begin{equation}
\Sigma_{\rm CO(3-2)} \equiv P_{\rm CO(3-2)}\,A_{\rm em}^{-1}\;,
\end{equation}
where $A_{\rm em}$ is the galaxy surface area (when looked face-on)
from which the CO(3--2) line is emitted. This area
is calculated by $A_{\rm em} = l_{\rm x}^2$ for the edge-on galaxies
of our sample ($i > 60^{\circ}$), and by
$A_{\rm em} = l_{\rm x}\,l_{\rm y}$ for the other objects,
to account for the different orientations.

We find that the two starburst galaxies NGC\,253 and M\,82 show
the highest value of $\Sigma_{\rm CO(3-2)}$, whereas most other starburst
galaxies (including NGC\,2146) show values a factor 3 smaller. M\,83 can
be found somewhere in between. The spiral galaxies NGC\,891 and especially
M\,51 are another considerable factor weaker, despite the fact that only
the extent of the central emission peak is used to calculate
$\Sigma_{\rm CO(3-2)}$ in both cases. At the lower end of
this scale we find NGC\,278, in agreement with e.g.\ infrared luminosity or
\ion{H}{i} mass, which point -- together with our other results on this
object -- to NGC\,278 being a rather small and quiet
galaxy. The ``mild starburst'' NGC\,4631 has an even lower value
for $\Sigma_{\rm CO(3-2)}$ than NGC\,278, reflecting also the large size
of the area from where CO(3--2) emission is detected.

\subsection{Line ratios and excitation}

By comparing the measured velocity-integrated intensities $I_{\rm CO(3-2)}$
with data from the (1--0) and (2--1) transitions, we can calculate line
ratios by
\begin{equation}
R_{3,1} = I_{\rm CO(3-2)}/I_{\rm CO(1-0)}
\end{equation}
(and $R_{3,2}$ analog). We have made fully sampled maps of the galaxies;
in addition the spatial resolution of our observations allows us to
determine line ratios for the disks and the nuclei of the galaxies
separately.

The line ratios we obtained are given in Table \ref{tab:results}. Care
must be taken when the data for the various transitions are observed
with different resolutions: in this case the process of calculating
line ratios introduces a larger error (e.g.\ for M\,83).
In addition, many authors don't give error estimates at all. Thus the
errors given in Table \ref{tab:results} mainly reflect the errors in
the measured CO(3--2) intensities and only relative errors in the used
CO(1--0) data sets. However, any systematical and calibration errors
in the used CO(1--0) data affect both the centre and disk values
for the line ratios.

The values we find for $R_{3,1}$ cover the range from 0.4 (in the centre of
NGC\,891) up to 1.4 (in the centre of NGC\,3628 and M\,83). In localized
regions within the galaxies the ratios may be even higher, but they cannot
be measured with limited spatial resolution. We should also note that the
observed values are distance-dependent, since our telescope beam
corresponds to different linear scales for the different galaxies.

When comparing the obtained line ratios with other paramters, we find
that there is no obvious trend with any other quantity. All galaxies
classified as starbursts show $R_{3,1} = 1.2 - 1.4$ in the centre,
except M\,82 and NGC\,253 with values of 1.0 and 0.8, respectively
(note, however, the uncertain calibration for NGC\,253 as mentioned
in subsection \ref{section:results_n253}). Two galaxies
show an exceptionally low ratio: NGC\,891 and M\,51. Both are usually
considered as relatively normal galaxies without a nuclear starburst, but
with ongoing star formation in the spiral arms.

\subsection{Summary and outlook}

We have obtained extended maps of the CO(3--2) emission in twelve nearby
galaxies, using the Heinrich-Hertz-Telescope on Mt.\ Graham, Arizona.
Our observations were not restricted to starburst galaxies, but include
also ``normal'' spiral galaxies. Further not only the central
positions of the objects have been observed, as in earlier studies. Hence
for the first time we were able to estimate spatially integrated parameters
of the CO(3--2) emission in the observed objects, to measure the
extent of the emission, and to investigate variations within these
objects.

We find that the CO(3--2) emission in the observed galaxies is not confined
to the nucleus. In some cases the CO(3--2) is as extended as the CO(1--0)
emission. Nevertheless, the CO(3--2) is more concentrated to star forming
structures (nuclear regions and spiral arms), which is shown by the decrease
of the CO(3--2)/CO(1--0) line ratios within the galaxies from the very centre
to regions further out.
The CO(3--2) luminosity is enhanced in objects which contain
a nuclear starburst or morphological peculiarities. The power emitted
in the CO(3--2) line from the central regions is highest in the starburst
galaxies NGC\,2146, M\,82, NGC\,3628, and in the spiral galaxy M\,51.
When normalized to the size of the emission region, the CO(3--2) emission
is a factor 3 higher in the starbursts NGC\,253 and M\,82 than in
most other objects.
With the present spatial resolution, the line ratios $R_{3,1}$ seem to be
independent of Hubble type, color, or luminosity. Most galaxies with
enhanced central star formation show line ratios of $R_{3,1} \sim 1.3$
in the very centre and $\sim 1.0$ at a radius of about 1\,kpc.
Those objects with a ring-like molecular gas distribution (NGC\,253, M\,82,
and NGC\,4631) show lower ratios.
The two galaxies that have CO(3--2) emission distributed over their spiral
arms (NGC\,891 and M\,51) show very low line ratios ($\sim 0.5$).
Further our data
suggest that even if starbursts are localized phenomena, they imply
different properties of the molecular gas also in the disks of the host
galaxies. This is shown by line ratios that are systematically higher in
the disks of starburst galaxies than in those of normal ones.

These observations have shown that it is necessary to study several
types of galaxies in order to obtain results which are not biased towards
starburst objects. Further it is not sufficient to observe only one point
per galaxy, since this cannot account for differences in the CO(3--2)
spatial distribution.

In order to improve this study, we will survey more galaxies in the future.
This will lead to better statistics of the findings presented here. Even
higher transitions of $^{12}$CO will be included, as well as observations
of other isotopomers to restrict interpretations concerning optical
depth effects.

\begin{acknowledgements}
We thank R. Beck, W. Huchtmeier, and R. Kothes for their collaboration
during the observations. Further we thank the SMTO staff and D.\ Muders
for their help at the telescope, and T.\ Wilson for his comments on the
manuscript. M.D. acknowledges financial support by a joint CNRS/MPG
stipendium, and C.N. acknowledges support by the Studien\-stiftung des
Deutschen Volkes.
\end{acknowledgements}

\end{document}